\documentclass{aa}

\usepackage{physics}
\usepackage{diagbox}
\usepackage{hyperref}
\newcommand{\diff}{\mathrm{d}}
\newcommand{\xir}{\xi_r}

\newcommand{\dis}{\Vec{\delta r}}
\usepackage{titlesec}

\titleformat{\paragraph}
{\normalfont\sffamily}{\theparagraph}{1em}{}
\titlespacing*{\paragraph}
{0pt}{3.25ex plus 1ex minus .2ex}{1.5ex plus .2ex}

\newcommand{\numax}{\nu_{\text{max}}}
\newcommand{\dnu}{\Delta\nu}
\newcommand{\operateur}[2]{\hat{\mathcal{#1}}_{#2}}
\newcommand\orc[1]{\href{https://orcid.org/#1}{\includegraphics[width=3mm]{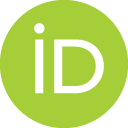}}}
\usepackage{graphicx}
\usepackage{subfigure}
\usepackage{txfonts}
\begin{document}

      \title{Near-degeneracy effects in Quadrupolar Mixed Modes}

   \subtitle{From an Asymptotic Description to Data Fitting}

   \author{B. Liagre
          \inst{1,2,3}\orc{0009-0008-7869-7430}
          \and
          A. Desai\inst{1}\orc{0009-0008-9877-5512}\and
          L. Einramhof\inst{1}\orc{0009-0002-5619-0598}
          \and L. Bugnet\inst{1}\orc{0000-0003-0142-4000}
          }

   \institute{   Institute of Science and Technology Austria (ISTA), Am Campus 1, 3400 Klosterneuburg, Austria
         \and
             Université Paris-Saclay, Université Paris Cité, CEA, CNRS, AIM, 91191, Gif-sur-Yvette, France\\
             \email{bastien.liagre@cea.fr}
        \and
            Instituto de Astrof\'isica de Canarias (IAC), E-38205 La Laguna, Tenerife, Spain
             }

   \date{Received 07/11/2025 ; accepted }

 
  \abstract{
   Dipolar ($\ell=1$) mixed modes revealed surprisingly weak differential rotation between the core and the envelope of evolved solar-like stars. Quadrupolar ($\ell=2$) mixed modes also contain information on the internal dynamics, but are very rarely characterised due to their low amplitude and the challenging identification of adjacent or overlapping rotationally split multiplets affected by near-degeneracy effects. We aim to extend broadly used asymptotic seismic diagnostics beyond $\ell=1$ mixed modes by developing an analogue asymptotic description of $\ell=2$ mixed modes, explicitly accounting for near-degeneracy effects that distort their rotational multiplets.
   We derive a new asymptotic formulation of near-degenerate mixed $\ell=2$ modes that describes off-diagonal terms representing the interaction between modes of adjacent radial orders. This formalism, expressed directly in the mixed-mode basis, provides analytical expressions for the near-degeneracy effects. We implement the formalism within a global Bayesian mode-fitting framework, for a direct fit of all $\ell=0,1,2$ modes in the power spectrum density.
   We are able to asymptotically model the asymmetric rotational splitting present in various radial orders of $\ell=2$ modes observed in young red giant stars without the need for any numerical stellar modelling. Applied to the \textsl{Kepler} target KIC~7341231, our formalism yields core and envelope rotation rates consistent with previous numerical modelling, while providing improved constraints from the global and model-independent approach. We also characterize the new target KIC~8179973, measuring its rotation rate and mixed-mode parameters for the first time.
   As our framework relies on a direct global fit, it allows for much better precision on the asteroseismic parameters and rotation rate estimates than standard methods, yielding better constraints for rotation inversions. We place the first observational constraints on the asymptotic $\ell=2$ mixed mode parameters ($\Delta\Pi_2$, $q_2$, and $\varepsilon_{{\rm g},2}$), paving the way towards the use of asymptotic seismology beyond $\ell=1$ mixed modes. }


   \maketitle
%

\section{Introduction}
Understanding the internal dynamics of evolved solar-like stars remains one of the major challenges in asteroseismology. Internal dynamical processes can be probed using oscillations that consist of acoustic pressure modes (p modes) and buoyancy-driven gravity modes (g modes). In evolved solar-like stars, p and g modes can couple across their respective cavities, giving rise to mixed modes \citep[\textit{e.g.}][]{1989nos..book.....U, 2016PASJ...68...91T}, first detected by \citet{2011Sci...332..205B}. Mixed-mode analyses have revealed the core and the envelope rotation rates of subgiant and red giant stars \citep[\textit{e.g.}][]{2011Sci...332..205B,2012ApJ...756...19D,2012A&A...548A..10M,2014A&A...564A..27D,2016ApJ...817...65D,2018A&A...616A..24G,2024A&A...688A.184L}. However, measured rotational gradients are substantially weaker than those predicted by hydrodynamical theoretical models of angular momentum evolution \citep[\textit{e.g.}][]{2012A&A...544L...4E,2013A&A...549A..74M,2013A&A...555A..54C,2017A&A...599A..18E,2019A&A...631L...6E}. 
This discrepancy indicates the presence of missing, angular momentum transport mechanisms in our representation of angular momentum transport in stars. Several processes have been proposed to enhance transport, including the meridional circulation \citep{1992A&A...265..115Z,1998A&A...334.1000M,2004A&A...425..229M,2009A&A...495..271D}, hydrodynamical instabilities \citep{1992A&A...265..115Z}, mixed modes \citep{2015A&A...579A..30B,2025A&A...699A.310B}, internal gravity waves \citep{1993A&A...279..431S,2014ApJ...796...17F,2017A&A...605A..31P,2025ApJ...983L..38R}, and magnetic processes such as transport by Alfvén waves and magnetic tension \citep[\textit{e.g.} ][]{1987MNRAS.226..123M,2007A&A...474..145Z,2021A&A...646A..19T,2022A&A...661A.119G}, magneto-rotational instabilities \citep{2024A&A...683A..12M}, magnetic webs \citep{2025ApJ...989L...4S}, or transport by the Tayler-Spruit dynamo \citep[\textit{e.g.}][]{1973MNRAS.161..365T, 2002A&A...381..923S, 2019MNRAS.485.3661F, 2023A&A...677A...6M,2025ApJ...988..195S}. To date, no observational constraints have unequivocally identified any overarching dominant physical mechanism responsible for angular momentum transport along stellar evolution. \\

To better constrain the efficiency of such newly proposed angular momentum transport mechanisms, more refined seismic diagnostics are required. So far, most efforts to constrain stellar internal rotation rates have focused on dipolar mixed modes, \textit{i.e.}, modes with degree $\ell=1$. Extending the analysis beyond dipole ($\ell=1$) modes to include quadrupole mixed modes ($\ell=2$) offers the prospect of significantly strengthening and adding constraints on internal dynamics. This extension not only increases the number of available observational probes but also provides sensitivity to different dynamical kernels, as modes of different degree sample distinct regions of the stellar interior, particularly within the convective zone \citep{2017A&ARv..25....1H}. The probing power of $\ell=2$ mixed modes and our ability to use them are limited by two major factors. (i) The p- and g-mode cavities sit further away from each other than those of $\ell=1$ modes, leading to an expected much weaker coupling strength reducing $\ell=2$ mixed modes height in the power spectrum density \citep[hereafter PSD,][]{2009A&A...506...57D,2014A&A...572A..11G,2018A&A...618A.109M}. (ii) The presence of near-degeneracy effects between adjacent multiplets distort the splitting structure \citep{1967MNRAS.136..293L,1992ApJ...394..670D,2012ApJ...756...19D,2014A&A...564A..27D,2017A&A...605A..75D,2018A&A...618A.109M,2022ApJ...940...18O,2024A&A...688A.184L,2025arXiv250926319A} and prevent straightforward identification and interpretation of the $\ell=2$ mixed-mode frequency pattern. \\

While the above-mentioned point (i) is a detectability limit, (ii) reduces our ability to identify modes. Standard first-order perturbation theory predicts symmetric splittings in slowly-rotating stars \citep[\textit{e.g.}][]{2010aste.book.....A,2013A&A...549A..75G,2014A&A...564A..27D,2018A&A...616A..24G}. 
However, if two unperturbed modes are close in frequency, their eigenfunctions are no longer independent. The resulting near-degeneracy introduces off-diagonal interaction terms in the oscillation equations, producing an observable asymmetry in the mixed-mode frequency pattern
\citep{1967MNRAS.136..293L,1992ApJ...394..670D,2017A&A...605A..75D,2018A&A...618A.109M,2022ApJ...940...18O,2024A&A...688A.184L,2025arXiv250926319A}. Despite several studies investigating near-degeneracy effects on oscillation frequencies \citep[\textit{e.g.}][]{2017A&A...605A..75D,2018A&A...618A.109M,2022ApJ...940...18O,2024A&A...688A.184L,2025arXiv250926319A} and efforts to include $\ell=2$ modes in seismic analysis \citep{2012ApJ...756...19D,2014A&A...564A..27D,2017A&A...605A..75D,2013ApJ...767..158B}, no asymptotic framework has been developed yet to explicitly compute the near-degeneracy interaction terms in the mixed-mode basis. For instance, \citet{2024A&A...688A.184L} performed an implicit treatment of these effects, while \citet{2022ApJ...940...18O} carried out the calculation in the $\pi$–$\gamma$ decoupled mode basis. In this work, we present a new asymptotic formalism for mixed quadrupole modes that explicitly incorporates near-degeneracy effects through non-diagonal near-degeneracy terms in the mixed-mode basis. This approach enables us to compute the near-degeneracy terms directly in the asymptotic limit and solve the associated quadratic eigenvalue problem. It offers two main advantages: (i) it preserves the mixed-mode basis, which simplifies the implementation for direct use in data analysis, similarly to the method extensively used to analyse $\ell=1$ mixed modes, and (ii) it provides explicit asymptotic estimates of the off-diagonal near-degeneracy terms, allowing for closed-form near-degeneracy corrections derived from quantum perturbation theory.\\

This article is structured as follows: in Section \ref{sec:theo} we lay the theoretical basis for this work and show the development leading to the asymptotic expression of the off-diagonal near-degeneracy terms. In Section \ref{sec:priors}, we introduce the global Bayesian framework and priors we use for simultaneously fitting our radial and $\ell=1,2$ mixed modes. Finally, we apply our new formalism and analysis pipeline in Section~\ref{sec:fit} to two early red giant stars observed by the NASA \textit{Kepler} mission \citep{2010Sci...327..977B}. We start with KIC~7341231, already characterized with numerical methods by \citet{2017A&A...605A..75D}, to demonstrate the robustness of our prescription. We then perform the global fit of mixed modes in the new candidate KIC 8179973, which shows strong near-degeneracy effects in the quadrupolar modes, and report the results, before concluding in Section~\ref{sec:ccl}.


\section{New developments on near degeneracy effects}\label{sec:theo}
\subsection{Theoretical context}
We begin by outlining the theoretical background and notation underlying our new asymptotic formulation. This framework is developed within the standard theory of adiabatic stellar oscillations.

\subsubsection{The general rotating eigenvalue problem}
As shown in \textit{\textit{e.g.}} \cite{2010aste.book.....A}, in the case of a non-rotating spherical star, the stellar adiabatic oscillation problem can be cast as an eigenvalue equation:  
\begin{equation}\label{eigenval_eq}
    \,\frac{\partial^2\dis}{\partial t^2} = \,\hat{\mathcal{L}}\,\dis,
\end{equation}  
where $\hat{\mathcal{L}}$ denotes the linear operator that encapsulates the perturbations of the equilibrium due to volumetric forces and acts on the displacement field $\dis$ associated with stellar oscillations:
\begin{equation}
\hat{\mathcal{L}}\,\dis
= \frac{\nabla p'}{\rho}
  + \nabla \Phi'
  - \frac{\rho'}{\rho^{2}} \, \nabla p \, .
\end{equation}
where $\rho$ and $p$ are the equilibrium pressure and density of the star and $p',\rho',\Phi'$ are the Eulerian perturbations in pressure, density and gravitational potential.\\
The solutions of this eigenvalue problem are given by a set of eigenvalues $\omega_{0,n,\ell,m}$ and eigenfunctions $\dis_{0,n,\ell,m}$ that can be decomposed on the spherical harmonics basis as:
\begin{equation}
    \begin{split} \dis_{0,n,\ell,m}&=\left[\xi_{0,r,\ell,n,m}(r)Y_\ell^m(\theta,\phi)\Vec{e}_r\right.\\&\left.+\xi_{0,h,\ell,n,m}(r)\left(\frac{\partial Y_\ell^m}{\partial \theta}\Vec{e_\theta}+\frac{1}{\sin(\theta)}\frac{\partial Y_\ell^m}{\partial\varphi}\Vec{e_\varphi}\right)\right]\exp(-i\omega t)\,.
    \end{split}
\end{equation} 
In this basis $n$ is the number of nodes of the eigenfunction along the radius, $\ell$, and $m$, the angular degree and azimuthal number of the mode.\\
In the absence of any perturbation that breaks the spherical symmetry of the initial problem, the modes with the same radial order $n$ and angular degree $\ell$ but different azimuthal order $m$ are degenerate in frequency, such that $\omega_{0,n,\ell} = \omega_{0,n,\ell,m}\,\forall m$. In the rest of the article, we will refer to those eigenfunctions as unperturbed eigenfunctions. We will index them by their $n$ only, as our physical framework does not allow coupling between modes of different degrees $\ell$. Hence, the set of unperturbed eigenfunctions will be the set of all $\dis_{0,i}=\dis_{0,i,\ell,m}$ where $i$ plays the role of $n$ such that:
\begin{equation}
    \begin{split} \dis_{0,i}&=\left[\xi_{0,r,i}(r)Y_\ell^m(\theta,\phi)\Vec{e}_r\right.\\&\left.+\xi_{0,h,i}(r)\left(\frac{\partial Y_\ell^m}{\partial \theta}\Vec{e_\theta}+\frac{1}{\sin(\theta)}\frac{\partial Y_\ell^m}{\partial\varphi}\Vec{e_\varphi}\right)\right]\exp(-i\omega t)\,.
    \end{split}
\end{equation} 
In this equation, the $r$ index refers to the part of the eigenfunction that oscillates vertically along the radius, while the $h$ index refers to the horizontally oscillating part of the eigenfunction.
\\

When including stellar rotation into the motion equation, the spherical-symmetry-induced degeneracy is lifted and the solutions to the eigenvalue problem depend on $m$. At leading order in $\Omega$, the angular rotation rate of the star, the eigenvalue problem becomes \citep[\textit{e.g.}][]{2010aste.book.....A,2017A&A...605A..75D,2025arXiv250926319A}:  
\begin{equation}\label{eigenval_pert}
\begin{split}
    -\omega^2\,\dis
    + 2\omega\left(m\Omega - i\,\Vec{\Omega}\times\right)\dis
    &= \left(-\omega^2 + \omega\hat{\mathcal{L}}_{\mathrm{rot},m}\right)\dis \\
    &= \hat{\mathcal{L}}\,\dis,
\end{split}
\end{equation}  
where we assume $\Vec{\Omega}=\Omega(r,\theta)\left(\cos(\theta)\Vec{e}_r-\sin(\theta)\Vec{e}_\theta\right)$ and
where $\dis$ is the perturbed eigenfunction.\\
We then decompose $\dis$ on the basis of unperturbed eigenmodes such as:
\begin{equation}\label{eq:decomp}
    \frac{\dis}{\sqrt{I}} = \sum_i \frac{a_i}{\sqrt{I_{0,i}}}\,\dis_{0,i},
\end{equation}  
with
\begin{equation}
    I = \langle \dis,\dis\rangle\, ,
\end{equation}
where $\langle\cdot,\cdot\rangle$ is the inner product defined as (for two vectors $\Vec{a}$ and $\Vec{b}$ and $\rho$ the local density of the star):
\begin{equation}
    \langle\Vec{a},\Vec{b}\rangle = \int_\mathrm{star}\rho\Vec{a}^*\Vec{b}\mathrm{d}^3r\,.
\end{equation}
Thus Eq. (\ref{eigenval_pert}) can be written in a matrix form:
\begin{equation}
    \label{eq:mat_eigval}\omega^2\mathbb{I}\Vec{a}-\omega \mathbf{R}\Vec{a}-\mathbf{L}\Vec{a}=0\,,
\end{equation}
where $\mathbb{I}$ is the identity matrix, $\mathbf{L}=\mathrm{diag}(\omega_{0,i}^2)$, $\Vec{a}$ is the vector of the $a_i$ (\textit{i.e.} the coefficients of the decomposition of $\dis$ in the unperturbed base following Eq. (\ref{eq:decomp})) and:
\begin{equation}
    R_{i,j}=\frac{\langle\dis_{0,i},\operateur{L}{\mathrm{rot},m}\dis_{0,j}\rangle}{\sqrt{I_{0,i}I_{0,j}}}\quad.
\end{equation}
Assuming a rotational profile that only depends on the radius of the star $\Omega=\Omega(r)$, the matrix elements of $R$ can be written \citep[\textit{e.g.}][]{2017A&A...605A..75D}:
\begin{equation}
\left<\dis_{0,i},\operateur{L}{\mathrm{rot},m}\dis_{0,j}\right>=2m\int_\mathrm{star}K_{i,j}(r)\Omega(r)\diff r
\end{equation}
with 
\begin{equation}
\begin{split}
    &K_{i,j}(r)=\rho r^2\times\\&\left(\xi_{0,r,i}^*\xi_{0,r,j}+(L^2-1)\xi_{0,h,i}^*\xi_{0,h,j}-\xi_{0,h,i}^*\xi_{0,r,j}-\xi_{0,r,i}^*\xi_{0,h,j}\right)\,,
    \end{split}
\end{equation}
where we have introduced the notation $L=\sqrt{\ell(\ell+1)}$, and where $\dis_{0,i}$ and $\dis_{0,j}$ are the eigenfunctions of two mixed modes unperturbed by rotation.

\subsubsection{The Two-zone rotating model}

In the rest of the developments, we use the general expression for the matrix elements $R_{ij}$ and discuss the simplifications that arise under the assumption of a two-zone rotation model. As is customary in the literature \citep[\textit{e.g.}][]{2012A&A...548A..10M,2018A&A...616A..24G}, we now consider a simplified model of rotation with two zones - the core \textit{i.e.} the g-mode cavity and the envelope \textit{i.e.} the p-mode cavity - rotating at fixed rotational rates $\Omega_\mathrm{core}$ and $\Omega_\mathrm{env}$ respectively.
\\

\subsection{First order symmetric rotational effects: diagonal terms}

At first perturbative order (that is to say, neglecting the coupling elements between eigenmodes of different $n$), an eigenfrequency will symmetrically lift its degeneracy and create what we will refer to as a rotational splitting of width  \citep[\textit{e.g.}][]{2010aste.book.....A,1989nos..book.....U}:

\begin{equation}
    \label{splitting_int}\omega_{i}-\omega_{0,i} = \delta\omega_i = \frac{1}{2I}\left<\dis_{0,i},\operateur{L}{\mathrm{rot},m}\dis_{0,i}\right>\quad.
\end{equation}
\\
The modes of the same $\ell$ and different $m$ are no longer at the same frequency but are at first order distributed symmetrically around the peak of $m = 0$. \\
The coupling element:
 \begin{equation}
     \frac{1}{I}\langle\dis_{0,i},\operateur{L}{\mathrm{rot},m}\dis_{0,i}\rangle
 \end{equation}
 may be asymptotically expressed using the seismic $\zeta$ function.
For a given unperturbed mixed mode, the asymptotic expression of the \(\zeta\)-function provides the ratio between the mode inertia in the g-mode cavity and the total mode inertia \citep[][see also section \ref{sec:mixed_modes} for the asymptotic expression of this quantity]{2013A&A...549A..75G}:
\begin{equation}
    \zeta(\omega) = \frac{I_{{\rm g}}}{I}\,.
    \label{eq:zetaI}
\end{equation}

Based on this relation, the asymptotic expressions for the diagonal elements of the coupling matrix within the two-zone model were shown to be given by \citep{2013A&A...549A..75G,2014A&A...564A..27D}:
\begin{equation}\label{split_first_order}
\begin{split}
&\frac{\left<\dis_{0,i},\operateur{L}{\mathrm{rot},m}\dis_{0,i}\right>}{I_{0,i}}=2m\times\\&\left(\Omega_\mathrm{core}\zeta(\omega_{0,i})\left(1-\frac{1}{L^2}\right)+\Omega_\mathrm{env}(1-\zeta(\omega_{0,i}))\right)\quad.
\end{split}
\end{equation}
We will use this description, excluding near-degeneracy effects (\textit{i.e.} off-diagonal near-degeneracy terms) to fit the $\ell=1$ modes in Sec. \ref{sec:fit} because the targets we choose have a large frequency separation $\dnu>20$ where we expect little to no influence of near-degeneracy effects \citep{2025arXiv250926319A}.
In the case of more evolved stars in the regime of low asymmetry splittings (typically for $11<\dnu<13$ depending on the core rotation rate according to \citet{2025arXiv250926319A}), our framework provides an explicit second order perturbative correction to the rotational splitting (based on Eq.~(\ref{off-diag-factor}), see Appendix \ref{pert_treat} for more details).\\
This approach eliminates the need to solve the implicit equation used by \cite{2024A&A...688A.184L} to reproduce the rotational asymmetries in $\ell=1$ mode splittings.

\subsection{Near-degeneracy effects: non-diagonal terms}
This symmetry in the splitting of the modes can be broken when the non-diagonal elements (\textit{i.e.} $i\neq j$) of the matrix $R$ become non-negligible compared to the diagonal. This is expected to happen when two modes have a frequency separation comparable to the rotation rate of the star \textit{i.e.} when $\lvert\omega_i-\omega_j\rvert\sim\Omega$ \citep{1992ApJ...394..670D,2017A&A...605A..75D,2025arXiv250926319A}.
\\
The off-diagonal terms, which are those responsible for the near-degeneracy asymmetry effects, can be written as \citep{2017A&A...605A..75D}:
\begin{equation}\label{eq:full_coupling_term}
\frac{\left<\dis_{0,i},\operateur{L}{\mathrm{rot},m}\dis_{0,j}\right>}{\sqrt{I_{0,i}I_{0,j}}}=2m\left(\gamma_{c,ij}\Omega_\mathrm{core}+\gamma_{e,ij}\Omega_\mathrm{env}\right)\,,
\end{equation}
with
\begin{equation}
\gamma_{c,ij}=\frac{1}{\sqrt{I_{0,i}I_{0,j}}}\int_\mathrm{g
}K_{i,j}(r)\diff r\,,
\end{equation}
and 
\begin{equation}
    \gamma_{e,ij}=\frac{1}{\sqrt{I_{0,i}I_{0,j}}}\int_\mathrm{p}K_{i,j}(r)\diff r\;.
\end{equation}

Because the mixed-mode basis is orthogonal and that the oscillation verify $\xir\ll\xi_h$ (resp. $\xir\gg\xi_h$) in the g-mode cavity (resp. p-mode cavity) \citep[\textit{e.g.}][]{2010aste.book.....A,2021A&A...650A..53B,2021A&A...647A.122M}, \cite{2017A&A...605A..75D} shows that:
\begin{equation}
    \label{gamma}\gamma_{c,ij}=\frac{1}{\sqrt{I_{0,i}I_{0,j}}}\int_\mathrm{g}\rho r^2(L^2-1)\xi_{0,h,i}^*\xi_{0,h,j}\diff r\,, 
\end{equation}
and
\begin{equation}
    \gamma_{e,ij}=\frac{L^2}{1-L^2}\gamma_{c,ij}\;.
\end{equation}
 Hence in order to describe the off-diagonal terms of the eigenvalue problem it is sufficient to evaluate $\gamma_{c,ij}$.

\subsubsection{New asymptotic formula for the near degeneracy effects}
\label{asymp_coup}


In order to find an asymptotic formula for $\gamma_{c,ij}$ that could be applied to fit observations, we used the asymptotic JWKB (Jeffreys-Wentzel–Kramers–Brillouin) \citep[\textit{e.g.}][]{https://doi.org/10.1112/plms/s2-23.1.428,1926ZPhy...38..518W} form of the horizontal eigenfunctions in the g-mode cavity as given in \textit{e.g.} \cite{2021A&A...647A.122M}:
\begin{align}
    &\label{osc_gamm_h}\xi_h=-\frac{A}{\sqrt{L}\omega}\rho^{-1/2}r^{-3/2}\left(\frac{N^2}{\omega^2}\right)^{1/4}\sin\left(L\int_{0}^r\frac{N}{\omega}\frac{\diff r'}{r'}-\frac{\pi}{4}\right),
\end{align}
Where $A$ is the amplitude of the mode and $N$ the Brunt–Väisälä frequency. This yields:
\begin{equation}
\begin{split}
    &\gamma_{c,ij} =\frac{1}{\sqrt{I_{0,i}I_{0,j}}}\int_\mathrm{g}\frac{A_{0,i,g}A_{0,j,g}(L^2-1)}{L\sqrt{\omega_{0,i}^3\omega_{0,j}^3}}\frac{N}{r}\times\\&\sin\left(L\int_{0}^r\frac{N}{r'}\diff r'\frac{1}{\omega_{0,i}}-\frac{\pi}{4}\right)\sin\left(L\int_{0}^r\frac{N}{r'}\diff r'\frac{1}{\omega_{0,j}}-\frac{\pi}{4}\right)\diff r.
\end{split}
\end{equation}
Asymptotically, the inertia of the mode in the g-mode cavity can be written as \citep[\textit{e.g.}][]{2017A&ARv..25....1H}
\[
I_g = \frac{A_g^2 \pi^2 L}{\omega^3 \Delta\Pi}
\quad \Leftrightarrow \quad
A_g = \sqrt{\frac{\Delta\Pi\, \omega^3 I_g}{L \pi^2}}\,,
\]
with $\Delta\Pi=\sqrt{\ell(\ell+1)}\Delta\Pi_\ell$ the reduced period spacing of the g modes. This yields the following result:

\begin{equation}
\begin{split}
    \gamma_{c,ij} =& \frac{1}{\sqrt{I_{0,i}I_{0,j}}}\int_\mathrm{g~cavity}\frac{\Delta\Pi(L^2-1)}{L^2\pi^2}\sqrt{I_{0,g,i}I_{0,g,j}}\frac{N}{2r}\times\\&\left[\cos\left(L\int_0^r\frac{N}{r'}\diff r'\frac{\omega_{0,j}-\omega_{0,i}}{\omega_{0,i}\omega_{0,j}}\right)\right.\\&\left.-\sin\left(L\int_0^r\frac{N}{r'}\diff r'\frac{\omega_{0,j}+\omega_{0,i}}{\omega_{0,i}\omega_{0,j}}\right)\right]\diff r \, .
\end{split}
\end{equation}
Where we have used the trigonometric identity $\sin(a)\sin(b)=\frac{1}{2}\left(\cos(a-b)-\cos(a+b)\right)$.
Using the change of variable\footnote{it is possible because in the g-mode cavity $\frac{N}{r}>0$ hence the integral is a strictly increasing function of $r$. This change of variables leads to $\diff s=\frac{N}{r}\diff r$}
\begin{equation}
s=\int_0^r\frac{N}{r'}\diff r'\,,
\end{equation}
and using the seismic $\zeta$ function, leads to the following expression:
\begin{equation}
\begin{split}
    &\gamma_{c,ij}=\frac{\Delta\Pi(L^2-1)}{2\pi^2L^2}\sqrt{\zeta_{0,i}\zeta_{0,j}}\times\\&\int_0^{\frac{2\pi^2}{\Delta\Pi}}\cos\left(Ls\frac{\omega_{0,i}-\omega_{0,j}}{\omega_{0,i}\omega_{0,j}}\right)-\sin\left(Ls\frac{\omega_{0,i}+\omega_{0,j}}{\omega_{0,i}\omega_{0,j}}\right)\diff s \, ,
\end{split}
\end{equation}
where we have used the fact that by definition of $\Delta\Pi$, $s(r_\mathrm{core})={2\pi^2}/{\Delta\Pi}$.
This is integrated to:
\begin{equation}\label{off-diag-factor}
\begin{split}
    \gamma_{c,ij} =& \frac{\Delta\Pi(L^2-1)}{\pi L^3}\sqrt{\zeta_{0,i}\zeta_{0,j}}\times\left[\frac{\nu_{0,i}\nu_{0,j}}{\nu_{0,j}-\nu_{0,i}}\sin\left(\frac{\pi L}{\Delta\Pi}\frac{\nu_{0,j}-\nu_{0,i}}{\nu_{0,j}\nu_{0,i}}\right)\right.
    \\&\left.+\frac{\nu_{0,i}\nu_{0,j}}{\nu_{0,j}+\nu_{0,i}}\left(\cos\left(\frac{\pi L}{\Delta\Pi}\frac{\nu_{0,j}+\nu_{0,i}}{\nu_{0,j}\nu_{0,i}}\right)-1\right)\right]\, ,
\end{split}    
\end{equation}
where we have expressed the quantities as a function of the frequency $\nu={\omega}/{2\pi}$ for ease of implementation. \\

An important point to highlight is that the quantity $\gamma_{c,ij}$, which we derived in this work as part of a new asymptotic formulation including off-diagonal rotational near-degeneracy terms, is expressed using the exact same asymptotic parameters traditionally used in first-order symmetric splitting approaches \citep[\textit{i.e.} in the characterization of mixed $\ell=2$ modes, $\Delta\Pi_2$ replaces $\Delta\Pi_1$ and $\zeta_{2}$ replaces $\zeta$ from the frameworks of][]{2012A&A...548A..10M,2013A&A...549A..75G,2014A&A...564A..27D}. 
This shows that we can achieve a better degree of precision on core and envelope rotation measurements by including additional constraints from $\ell=2$ modes without the need to change the usual formalism and analysis tools.\\
\begin{figure}
    \centering
    \includegraphics[width=0.9\linewidth]{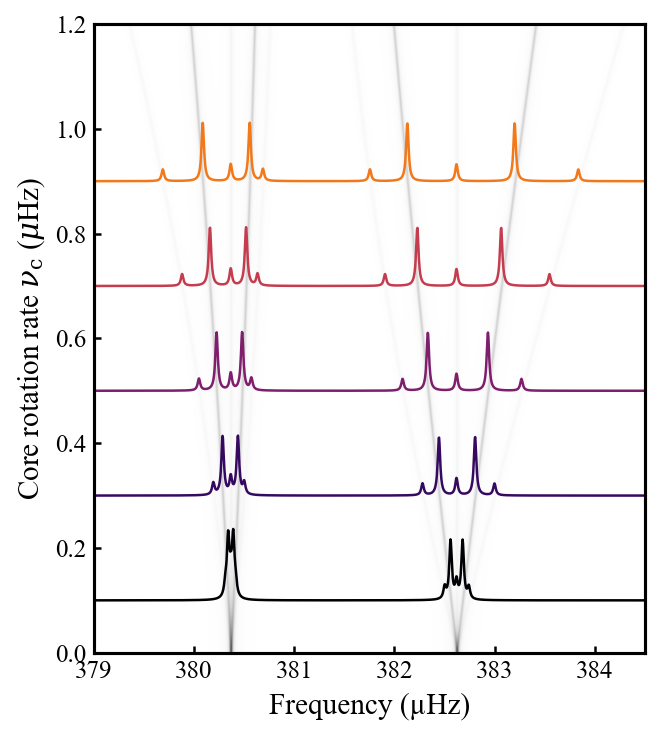}
    \caption{Illustration of the effect of near degeneracy on two coupled $\ell=2$ multiplets as the core rotation rate $\nu_\mathrm{c}=\Omega_\mathrm{core}/2\pi$ increases, for a fixed inclination angle of $i=42^\circ$. Each horizontal modelled PSD shows the combined visibility profile of the ten components (two sets of $m=-2,-1, 0, +1,+2$ modes) at a given $\nu_\mathrm{c}$. The greyscale background indicates how the multiplet structure evolves continuously with rotation. All the frequencies were computed by solving Eq. (\ref{eq:mat_eigval}) with off-diagonal terms provided by the asymptotic formulation of Eq. (\ref{off-diag-factor}) and diagonal terms given by Eq. (\ref{split_first_order}). This example highlights how near-degenerate interactions modify the usual pattern of rotationally split multiplets and generate asymmetries in the frequency pattern. }
    \label{fig:nd_eff}
\end{figure}

Our asymptotic formulation in Eq.~(\ref{off-diag-factor}) can be compared with the derivation of \cite{2022ApJ...940...18O} in the $\pi -\gamma$ decoupled mode basis.
When the argument of the $\sin$ in $\gamma_{c,ij}$ is small, \textit{i.e.}:
\begin{equation}
\frac{\pi L}{\Delta\Pi}\frac{\nu_j-\nu_i}{\nu_j\nu_i}=\pi\frac{P_i-P_j}{\Delta\Pi_\ell}\ll1 \, ,
\end{equation}
where $P_i$ (resp. $P_j$) is the period of the mode i (resp j), Eq.~(\ref{off-diag-factor}) becomes equivalent to $\gamma_c\approx\frac{L^2-1}{L^2}\sqrt{\zeta_i\zeta_j}$ using a first-order Taylor expansion. This gives $\gamma_e\approx-\sqrt{\zeta_i\zeta_j}$. This last approximation yields a result that echoes equation (35) of \cite{2022ApJ...940...18O}. This situation is most frequent around p-dominated mixed modes. Indeed for two consecutive mixed modes:
\begin{equation}
    \frac{\pi L}{\Delta\Pi}\frac{\nu_j-\nu_i}{\nu_j\nu_i}=\frac{\pi}{\Delta\Pi_\ell}\left(\frac{1}{\nu_i}-\frac{1}{\nu_j}\right)\sim\pi\langle\zeta\rangle \, ,
\end{equation}
where: 
\begin{equation}
    \langle\zeta\rangle=\left(\frac{1}{\nu_i}-\frac{1}{\nu_j}\right)\left(\int_{\frac{1}{\nu_j}}^{\frac{1}{\nu_i}} \frac{\mathrm{d}P}{\zeta(P)}\right)^{-1}
\end{equation}
with $P={1}/{\nu}$ following \cite{2015A&A...584A..50M} (see their Eq. (12)).
Hence, for a small $\langle\zeta\rangle$ (or equivalently a very p-dominated couple of modes), the quantity $\pi\langle\zeta\rangle$ will be small.
On the contrary, in the far resonance limit when 
\begin{equation}
    \frac{\pi L}{\Delta\Pi}\frac{\nu_j-\nu_i}{\nu_j\nu_i}\gg1\, ,
\end{equation}
  the non-diagonal terms become negligible compared to the diagonal ones and we recover symmetrical splittings.

\subsubsection{Validation of the new asymptotic formula for near degeneracy effects}\label{sec:dh_test}

To further verify the validity of our new asymptotic formulation of near-degeneracy effects, we sought to reproduce the results of \citet{2017A&A...605A..75D}. For that, we use exactly the same input parameters adopted in their study, only letting free the core and envelope rotation rates (Eq.~\ref{eq:full_coupling_term}). \\

KIC 7341231 is, to date, the only star in which the near-degeneracy effects of $\ell=2$ modes have been characterised in detail in one radial order. In their analysis, \citet{2017A&A...605A..75D} adopted a different approach from ours: they computed a one-dimensional stellar model using \texttt{Cesam2k} \citep{1997A&AS..124..597M}, assuming a two-zone rotation profile. They then explicitly computed $\gamma_c$ from the rotational kernels and solved the full eigenvalue problem. In contrast, our work relies on a fully parametrised asymptotic formulation of $\gamma_c$, which allows us to directly compare our results with those obtained from their numerical approach. Specifically, we fitted the core and envelope rotation rates of the two-zone model based on the $\ell=2, m=\pm2$ rotational splittings obtained from their peakbagging analysis of two $\ell=2$ multiplets $a$ and $b$ (see their Table~1), together with the $m=0$ mode frequencies ($\nu_{0,a}$ and $\nu_{0,b}$) and the corresponding trapping fractions $\zeta_a$ and $\zeta_b$ reported in their Table~3. To compute the $m=\pm2$ components of each mixed  $a$ and $b$, we evaluated the eigenvalues of the following matrix, as done by \citet{2017A&A...605A..75D} in their Appendix~A:
\begin{equation}
\small
    \begin{bmatrix}
        \nu_{0,a}^2 + 2\nu_{0,a}\dfrac{\left<\dis_{0,a},\operateur{L}{\mathrm{rot},m}\dis_{0,a}\right>}{2\pi I_{0,a}} &
        2\nu_{0,a}\dfrac{\left<\dis_{0,a},\operateur{L}{\mathrm{rot},m}\dis_{0,b}\right>}{2\pi\sqrt{I_{0,a}I_{0,b}}} \\[8pt]
        2\nu_{0,b}\dfrac{\left<\dis_{0,b},\operateur{L}{\mathrm{rot},m}\dis_{0,a}\right>}{2\pi\sqrt{I_{0,a}I_{0,b}}} &
        \nu_{0,b}^2 + 2\nu_{0,b}\dfrac{\left<\dis_{0,b},\operateur{L}{\mathrm{rot},m}\dis_{0,b}\right>}{2\pi I_{0,b}}
    \end{bmatrix}.
    \label{eq:degeneracy_matrix}
\end{equation}
This formulation is an approximation of (\ref{eq:mat_eigval})  introduced by \citet{2017A&A...605A..75D}. Since they use this approximated form in their article, we choose to do the same when vetting our results in order to stay as close to the article as possible. This is the only part of the article in which we compute the splittings due to near-degeneracy effects with this approximation, in the rest of the article we solve the full quadratic eigenvalue problem (\ref{eq:mat_eigval}).\\
In our formulation, the off-diagonal elements of this matrix were computed using Eq.~(\ref{off-diag-factor}), while the diagonal terms were derived from Eq.~(\ref{split_first_order}) \citep{2013A&A...549A..75G,2014A&A...564A..27D}.
\\

The values and uncertainties on the inferred core and envelope rotation rates were estimated using a Monte Carlo approach. The resulting values are shown in Fig.~\ref{fig:reproduction_Dh}. Using identical mixed-mode parameters but replacing their numerical model with our asymptotic formulation, we achieved an excellent agreement with the published results. We obtained $\Omega_\mathrm{core}/2\pi = 772 \pm 9~\mathrm{nHz}$ and $\Omega_\mathrm{env}/2\pi = 45 \pm 6~\mathrm{nHz}$, while \cite{2017A&A...605A..75D} reported $\Omega_\mathrm{core}/2\pi = 771 \pm 13~\mathrm{nHz}$ and $\Omega_\mathrm{env}/2\pi = 45 \pm 12~\mathrm{nHz}$. The close agreement between both methods confirms that the asymptotic formulation provides a robust and accurate description of the observed rotational splittings and can be used to analyse asteroseismic observations of $\ell=2$ modes without the need of numerical models. We also hint a better precision on our fitted measurement, which will be confirmed in Section~\ref{sec:fit}.

\section{Data analysis framework}\label{sec:priors}
We now aim to apply the asymptotic formalism described by Eq.~(\ref{eq:full_coupling_term}), (\ref{off-diag-factor}) and (\ref{split_first_order}) to perform global fits of the power spectral density (PSD), thereby constraining mixed modes parameters and core and envelope rotation rates using all available data. In particular, we deliberately avoid performing peakbagging prior to model fitting, as it can often be ambiguous whether a given peak represents a genuine oscillation mode or merely stochastic noise. The main advantage of fitting the PSD directly with a physically informed model based on asymptotic relations is that the distinction between signal and noise naturally emerges from the fit itself, without requiring any subjective preselection of peaks.  

To accomplish this, we employ a Monte-Carlo Markov Chain (MCMC) approach within a Bayesian framework to fit the oscillation modes in the PSD. The p- and mixed-mode patterns are created using the \texttt{sloscillations} code developed in \citet{2019MNRAS.488..572K,2023ApJ...954..152K}, extended to incorporate the near-degeneracy effects based on Section~\ref{sec:theo}. The synthetic PSD is generated following Sec. \ref{sec:synth_PSD} below.
\subsection{Synthetic PSD}\label{sec:synth_PSD}

\subsubsection{Description of the set-up of the asymptotic model}
The asymptotic model used in the fitting procedure predicts the full oscillation spectrum from a set of physically motivated parameters. For each angular degree $\ell~\in~\{0,1,2\}$, the model generates the underlying p- and g-mode frequencies from the asymptotic relations described below, which are then coupled through the mixed-mode formalism and corrected for rotational splitting at first order (for $\ell=1$) or through the full quadratic eigenvalue problem (for $\ell=2$). \\
The free parameters of the model therefore include the global p-mode pattern parameters ($\Delta\nu$, $\varepsilon$, $\delta\nu_{0,\ell}$, $\alpha$, $\beta_{0,\ell}$), the g-mode period-spacing and phase parameters ($\Delta\Pi_\ell$, $\varepsilon_{g,\ell}$), the p--g coupling strengths ($q_\ell$), and the core and envelope rotation rates ($\Omega_{\rm core}$, $\Omega_{\rm env}$), as well as the stellar inclination angle $\theta_{\rm inc}$. 
\\Each predicted mode is assigned a Lorentzian profile with analytically prescribed linewidth and height, and these components are summed together with a synthetic granulation and white-noise background to produce a fully specified model of the PSD.

\subsubsection{Pure pressure and gravity modes}
The pure pressure modes are computed using the following equation \citep[\textit{e.g.}][]{1980ApJS...43..469T,1990Natur.345..322E,2011A&A...525L...9M,2017ApJ...835..172L,2022A&A...663A.118B}:
\begin{equation}\label{asymp_p_gen}
\begin{split}
    &\nu_{{\rm p}_{n,\ell}}=\\&\left(n+\frac{\ell}{2}+\varepsilon+\beta_{0,\ell}(n-n_\mathrm{max})+\frac{\alpha}{2}(n-n_\mathrm{max})^2\right)\Delta\nu-\delta\nu_{0,\ell} \,,
\end{split}
\end{equation}
where $n_\mathrm{max}$ is the order of the mode of maximum power, $\varepsilon$ is the global p mode phase shift, $\delta\nu_{0,\ell}$ is the frequency shift of the modes from their first order asymptotic position \citep{1980ApJS...43..469T}, and $\alpha$ and $\beta_{0,\ell}$ are introduced ad hoc to capture the frequency dependence of $\Delta\nu$ and $\delta\nu_{0,\ell}$ \citep{1990Natur.345..322E,2017ApJ...835..172L,2022A&A...663A.118B}.
On the other hand, the pure gravity modes are computed using \citep[\textit{e.g.}][]{1980ApJS...43..469T}:
\begin{equation}\label{g_asymp}
    \nu_{{\rm g}_{n,\ell}}=\frac{1}{\Delta\Pi_\ell(n+\frac{1}{2}+\varepsilon_{\rm g})}\,,
\end{equation}
where $\varepsilon_{\rm g}$ is the phase shift of gravity modes.
\subsubsection{Mixed modes with rotation}\label{sec:mixed_modes}
Unperturbed mixed mode frequencies are computed using equation \citep{1979PASJ...31...87S}:
\begin{equation}
    \label{asymp_mix}\text{cotan}\left(\Theta_{\rm g}\right)\tan(\Theta_{\rm p})=q_\ell
\end{equation}
with \citep{2023ApJ...946...92O}:
\begin{equation}\label{phases}\begin{split}
    &\Theta_{\rm p} = \frac{\pi}{\Delta\nu}\left(\nu-\nu_{\rm p}\right),\\&\Theta_{\rm g} = \frac{\pi}{2}-\frac{\pi}{\Delta\Pi_\ell}\left(\frac{1}{\nu_{\rm g}}-\frac{1}{\nu}\right)\, ,
    \end{split}
\end{equation}
and where $q_\ell$ is the coupling factor between p- and g- modes for the corresponding order $\ell$ (we expect $q_2<q_1$ due to the shallower p-mode cavity for $\ell=2$ modes).\\
Then we introduce the effects of rotation on the unperturbed frequencies at first perturbative order for the $\ell=1$ components and solve the full rotating quadratic eigenvalue problem Eq.~(\ref{eq:mat_eigval}) with asymptotic matrix elements for $\ell=2$ (based on our asymptotic formulation Eq.~(\ref{off-diag-factor})) modes using the \texttt{slepc4py} \citep{10.1145/1089014.1089019,DALCIN20111124} Python library.\\

When computing the off diagonal elements of the rotation matrix, we use the asymptotic formula for $\zeta$ defined in Eq.~\ref{eq:zetaI} such that \citep{2013A&A...549A..75G}:
\begin{equation}
    \zeta(\nu)=\left[1+\frac{\nu^2\Delta\Pi_\ell}{\dnu}\frac{1}{q\cos^2(\Theta_p)+\frac{1}{q}\sin^2(\Theta_p)}\right]^{-1}\, ,
\end{equation}
using the above definition of $\Theta_p$.

\subsubsection{Modelling the linewidth, amplitudes and heights of the modes}
This paragraph describes the equations and methodologies used to compute the linewidths amplitudes and heights of the modes in the \texttt{sloscillations} code by \citet{2019MNRAS.488..572K} for the sake of completeness and clarity. These parameters are fully derived from scaling relations and analytical expressions and are thus not fitted parameters in the model. We adopt effective-temperature-dependent scaling relations for the p mode linewidths, parameterised as a function of temperature, following the empirical prescriptions of \citet{corsaro_asteroseismology_2012,2017ApJ...835..172L}. 
Radial-mode amplitudes follow a Gaussian envelope centred on $\nu_{\rm max}$ with width tied to the envelope full width at half maximum and a peak amplitude that scales with global parameters ($\nu_{\rm max}$, $\Delta\nu$). The per-degree amplitudes are set by standard photometric visibilities as in \cite{2019MNRAS.488..572K} and, for non-radial multiplets, by geometric inclination ($\theta_\mathrm{inc}$) factors \citep{2003ApJ...589.1009G}. For mixed $\ell = 1$ and $\ell = 2$ modes, we modulate the nominal p mode amplitudes by $(1-\zeta)^{\frac{1}{2}}$ \citep{2018A&A...618A.109M, Basu_Chaplin_2017}. Mode heights in the power spectrum are computed from the time-domain amplitudes and the linewidths using the finite-observation-length correction of \citet{Fletcher_2006}. For mixed modes, we use the mixed-mode generalisation that explicitly includes the inertia ratio $\zeta$ \citep{Basu_Chaplin_2017}, ensuring that g-dominated mixed modes appear with appropriately reduced height at fixed amplitude. Each mode contributes a Lorentzian profile to the PSD, centred at the asymptotic frequency and with the height and width described above. 
\\The background of the synthetic PSD is modelled with the usual super-Lorentzian granulation components and white noise \citep[e.g][]{2014A&A...570A..41K}. It is generated in the time domain via a Gaussian-process (GP) representation of convection and granulation following \cite{2019MNRAS.488..572K}, where a kernel encapsulates the covariance of stochastic surface motions (\textit{e.g.} \citealt{2017AJ....154..220F, 2019MNRAS.489.5764P, O’Sullivan_Aigrain_2024}). This data is Fourier transformed, and the resulting PSD provides the granulation floor plus white noise that forms the background in the PSD. On top of this background, the oscillation spectrum is added, producing a self-consistent synthetic PSD where both the convective noise and p-mode power originate from the same underlying time-domain process.
During the fitting procedure, the synthetic background is then normalised using the same procedure applied to the observed data (see Sec. \ref{sec:preprocess}).
\subsection{Description of the Bayesian fitting procedure}
\subsubsection{Priors on p modes and $\ell=1$ mixed modes}

Appendix~\ref{app:priors} describes the method used to define prior ranges for the rotational, p-mode, and $\ell=1$ mixed-mode parameters. These ranges are guided by both theoretical expectations and previous observational studies.

\subsubsection{Priors on the $\ell=2$ period spacing $\Delta\Pi_2$ and phase $\varepsilon_{g,2}$}\label{prior_l2}

Using the fact that the period spacing $\Delta\Pi=L\Delta\Pi_\ell$ is constant \citep{1980ApJS...43..469T}, we obtain 
\begin{equation}
\Delta\Pi_2=\frac{\Delta\Pi_1}{\sqrt{3}}\,.
\label{eq:dpi}
\end{equation}
Hence, once we have measured $\Delta\Pi_1$, we choose to assign a flat prior within $\pm5\%$ of the value of $\Delta\Pi_2$ from Eq.~(\ref{eq:dpi}).\\
As for $\varepsilon_{{\rm g},2}$, we have, to our knowledge, no theoretical grounds to constrain it. Hence, we chose to use an uninformative flat prior between 0 and 1, as done for $\varepsilon_{{\rm g},1}$ (see Appendix~\ref{app:priors}).

\subsubsection{Priors on the $\ell=2$ modes coupling parameter $q_2$}
We examine what constraints can be put on $q_2$, the coupling factor of $\ell=2$ modes, depending on the inclination angle of the star compared to the line of sight ($\theta_\mathrm{inc}$) (which we estimate by the preliminary fit of the $\ell=1$ modes --  Appendix \ref{app:priors}). When the $\ell=2,\, m=0$ modes are visible, we use the technique described in Sec. \ref{q_from_modes}. When the angle of inclination makes the $\ell=2$ and $m=0$ modes not visible, we use the result of Sec. \ref{up_lim_q2}.
\subsubsection{Priors on the $\ell=2$ modes coupling parameter $q_2$ using the $m=0$ modes}\label{q_from_modes}
Using equation (15) of \cite{2023ApJ...946...92O}, we obtain that for 2 consecutive mixed modes (with same $\ell$ order) of frequency $\nu_a$ and $\nu_b$:
\begin{equation}
\begin{split}    &\arctan\left(\frac{\tan\left(\Theta_{{\rm p},a}\right)}{q_\ell}\right)-\arctan\left(\frac{\tan\left(\Theta_{{\rm p},b}\right)}{q_\ell}\right)\\&=\frac{\pi}{\Delta\Pi_\ell}\left(\frac{1}{\nu_a}-\frac{1}{\nu_b}+\Delta\Pi_\ell\right)\quad.
\end{split}
\end{equation}
Using trigonometric relations, we show that this formula is equivalent to:
\begin{equation}\begin{split}
    &\frac{\tan\left(\Theta_{{\rm p},a}\right)-\tan\left(\Theta_{{\rm p},b}\right)}{q_\ell}\frac{1}{1+\frac{1}{q_\ell^2}\tan\left(\Theta_{{\rm p},a}\right)\tan\left(\Theta_{{\rm p},b}\right)}\\&=-\tan\left(\frac{\pi}{\Delta\Pi_\ell}\left(\frac{1}{\nu_b}-\frac{1}{\nu_a}\right)\right)\,,
    \end{split}
\end{equation}
which is conveniently a second-order polynomial in $1\over q_\ell$.\\
Hence, defining:
\begin{equation} 
\begin{split}
     K&=\tan\left(\frac{\pi}{\Delta\Pi_\ell}\left(\frac{1}{\nu_b}-\frac{1}{\nu_a}\right)\right)\\
    A&=K\tan\left(\Theta_{{\rm p},a}\right)\tan\left(\Theta_{{\rm p},b}\right)\\
B&=\tan\left(\Theta_{{\rm p},b}\right)-\tan\left(\Theta_{{\rm p},a}\right)\\
\Delta&=B^2-4AK\,,
\end{split}
\end{equation}
we obtain that $q_\ell$ is given by:
\begin{equation}
    \label{q_ana}q_\ell = \frac{2A}{B+\sqrt{\Delta}}\quad.
\end{equation}
In order to obtain $q_2$, we thus need $\Delta\Pi_2$, $\dnu$, $\nu_a,\nu_b,$ and the position of the $\ell=2$ p mode that is obtained from the radial mode frequency and $\delta\nu_{0,2}$.

\paragraph{Priors on the $\ell=2$ modes coupling parameter $q_2$ using the $\ell=1$ mode coupling parameter $q_1$}\label{up_lim_q2}
When the inclination angle of the rotation axis is such that the $m=0$ components of the multiplet splittings are not visible, we cannot use the above method to yield an estimate of $q_2$. Indeed the previous method relies on the frequencies of the pure mixed modes not affected by dynamical processes which is not the case for the components of a multiplet with $\lvert m\rvert>0$. Hence, we derive an upper bound of $q_2$ as a function of $q_1$. In order to do so,
let us use the approximation that in the evanescent zone $N\ll\omega\ll S_\ell$ where $N$ is the Brunt-Väisälä frequency and $S_\ell$ the Lamb frequency. In the approximation of a thick evanescent zone,  the transmission coefficient can be approximated as \citep{1989nos..book.....U,2016PASJ...68...91T}:
\begin{equation}
    t_{\ell,\mathrm{thick}}\approx\left(\frac{r_0}{r_\ell}\right)^L\,,
\end{equation}
where $r_0$ is the turning point between the g-mode cavity and the evanescent zone which is independent on $\ell$ and $r_\ell$ is the turning point at the limit between the evanescent zone and the p-cavity.\\
While this thick evanescent zone approximation may not be applicable for $\ell=1$ modes, we know from \cite{2016PASJ...68..109T} that $t_{\ell,\mathrm{thick}}>t_{\ell,\mathrm{thin}}$ where $t_{\ell,\mathrm{thin}}$ is the transmission coefficient computed with the asymptotic equation appropriate when the evanescent zone is thin \citep{2016PASJ...68..109T,2025A&A...700A...1V}. Hence, this approximation is appropriate in the search of an upper boundary on $q_2$. Because for the same frequency $r_2>r_1$ since $S_\ell$ is an increasing function of $\ell$, that implies that:
\begin{equation}
    t_2 \approx \left(\frac{r_0}{r_2}\right)^{\sqrt{6}}
    < \left(\frac{r_0}{r_1}\right)^{\sqrt{6}}
    \approx t_1^{\sqrt{3}} \quad .
\end{equation}
Using the fact that \citep{2016PASJ...68...91T}:
\begin{equation}
    q_\ell=\frac{1-\sqrt{1-t_\ell^2}}{1+\sqrt{1-t_\ell^2}}\, ,
\end{equation}
it finally comes that:
\begin{equation}
    \label{upper_bound_q2}
    q_2 < \frac{1 - \sqrt{1 - \left(\frac{4q_1}{(q_1+1)^2}\right)^{\sqrt{3}}}}
                 {1 + \sqrt{1 - \left(\frac{4q_1}{(q_1+1)^2}\right)^{\sqrt{3}}}} \; .
\end{equation}

\subsubsection{Likelihood adopted in the MCMC}
In the MCMC, we adopt the likelihood appropriate for a power spectrum whose noise follows a $\chi^2$ distribution with two degrees of freedom \citep{1984PhDT........34W,1994A&A...289..649T,2022A&A...663A.118B}. The corresponding log-likelihood is
\[
\ln \mathcal{L}
  = -\sum_i \left[ \frac{P_i}{M_i} + \ln M_i \right],
\]
where $P_i$ denotes the observed power and $M_i$ the model power at frequency bin $i$. The Bayesian inference procedure then adjusts the asymptotic parameters so as to maximise this likelihood, thereby ensuring that the resulting synthetic PSD matches the observed one as closely as possible.
The different components of the model are presented in the paragraphs below.

\subsubsection{Adopted fitting strategy}\label{Bayesfit}
In order to fit the asymptotic pattern to asteroseismic data, we used a three-step approach:
(i) In order to find priors on pressure mode parameters we run the global asteroseismic code PyA2Z \citep[Liagre et al. in prep, based on the A2Z code][]{2010A&A...511A..46M} which gives guesses of $\dnu$ the large frequency separation, $\numax$ the frequency of maximum power, and $\varepsilon$ the p mode phase shift (as well as $\Delta\Pi_1$ the period spacing of $\ell=1$ gravity modes).
    (ii) We fit the radial mode pattern and the first order split $\ell=1$ mixed-mode pattern following Sec.~\ref{prior_pl1} (\textit{i.e.} we fit $\dnu, \varepsilon, \delta\nu_{0,1}, \alpha,\Delta\Pi_1,q_1,\varepsilon_{g,1},\Omega_\mathrm{core}, \Omega_\mathrm{env},\beta_{0,1},$ and $\theta_\mathrm{inc}$ the inclination angle of the rotation axis compared to the line of sight for KIC 8179973) to get the parameters needed to obtain priors on $\ell=2$ modes from Eqs.~(\ref{eq:dpi}), (\ref{q_ana}), and (\ref{upper_bound_q2}). 
    (iii) Once we have the priors on all parameters (listed in Tab. \ref{tab:priors}), we perform the global asymptotic fit of the PSD ($\ell=0,1,2$ modes with rotation) following Appendix \ref{Prior_7341231} and \ref{Prior_8179973}.  We choose not to include $\ell\ge 3$ modes in our model, as their amplitude are very small.\\
    
While the mixed $\ell=1$ mode parameters and pressure mode parameters were already fitted in the first two steps of the procedure, we have chosen not to freeze them during the global fit so as to avoid biases in the inferred rotational and $l=2$ parameters. This resulted in a global fit, the cornerplot of which can be found in Appendix \ref{app:corner}.

\section{The first asymptotic fits of near-degeneracy effects in quadrupolar mixed modes}\label{sec:fit}

\subsection{Selection of candidates and data pre-processing}\label{sec:preprocess}
The first target used to validate this new asymptotic formulation of asymmetric rotational splitting is the one studied with a numerical modelling approach by \cite{2017A&A...605A..75D}: KIC 7341231. In order to find additional targets with mixed $\ell=2$ modes on which to apply our formulation of asymmetric rotational splittings, we chose to focus on stars in the early red giants/late subgiants evolutionary stage as they are expected to exhibit higher coupling factors for both $\ell=1$ and $\ell=2$ modes than more evolved stars. A convenient sample for exploring this evolutionary phase is the one from \citet{2025A&A...702A.144L} that re-analysed aliased asteroseismic data from $\approx 2000$ long cadence \textit{Kepler} lightcurves and found a new sample of $\approx 350$ young red giants with $\numax$ above or close to the Nyquist frequency of the \textit{Kepler} long cadence data. While the sample presented in \citet{2025A&A...702A.144L} contains more than 300 stars with pulsations above the Nyquist frequency and hence in the early red giant evolutionary stage, we wanted to analyse stars with a short-cadence timeseries to avoid misinterpreting aliased peaks as modes and with sufficient resolution and signal to noise ratio to work well with our fitting procedure. This significantly reduced the available number of targets, and among those, we found that KIC 8179973 was the most promising, as a visual inspection the echelle diagram of the star showed clear $\ell=2$ avoided crossings. We therefore selected KIC 8179973 as another good candidate for a global fit of $\ell=1,2$ mixed modes including near-degeneracy effects, as it exhibits mixed $\ell=2$ modes with an exceptional signal-to-noise ratio. \\

The seismic analyses were done using short-cadence \emph{Kepler} light curves filtered at 20 days obtained from the Mikulski Archive for Space Telescopes (MAST) archive. They were corrected with the \emph{Kepler} Asteroseismic data analysis and calibration Software \citep[\texttt{KADACS},][]{2011MNRAS.414L...6G},
to remove outliers, correct any jumps and drifts, and stitch together the quarters. All the observational gaps were interpolated using in-painting techniques based on a multi-scale discrete cosine transform \citep{2014A&A...568A..10G,2015A&A...574A..18P}.
 We isolate the oscillatory signal in the PSD from the granulation background, which dominates the PSD at frequencies around $\nu_{\rm max}$. We implement the background normalisation procedure using the empirical scaling relation established in \citet{2012A&A...537A..30M}:
\begin{equation}
    \mathrm{PSD \,(SNR)} = \frac{P}{(\nu_{\rm max}^{-2.41})\Big(\frac{\nu}{\nu_{\rm max}}\Big)^{-2.1}6.37\times10^6},
\end{equation}
\noindent
where $P$ represents either the observed or the synthetic power spectrum density. This normalization effectively flattens the granulation background while preserving the relative amplitudes of the oscillation modes. Subsequently, we isolate the region of interest around $\nu_{\rm max}$ (the extracted frequency window spans from $\nu_{\rm max} - 3\dnu \leq \nu_{\rm max} \leq \numax + 4 \dnu$).

\subsection{Results of the Bayesian asymptotic fitting}
\begin{figure}[h]
    \centering

    \begin{minipage}{\linewidth}
        \centering
        \includegraphics[width=\linewidth]{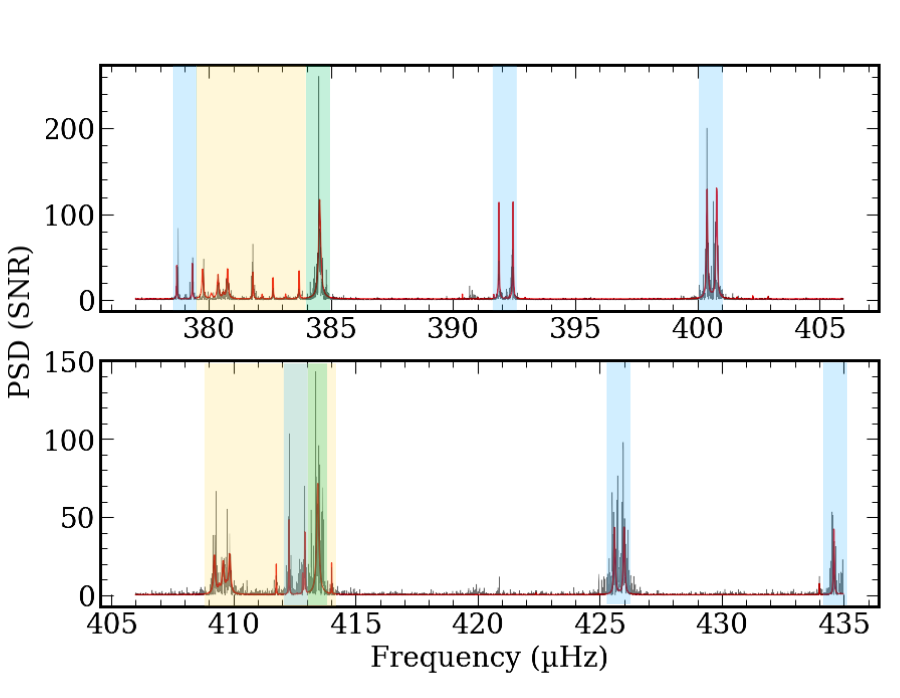}
        \textbf{(a)}
    \end{minipage}
    
    \vspace{0.5cm} 
    
    \begin{minipage}{\linewidth}
        \centering
        \includegraphics[width=\linewidth]{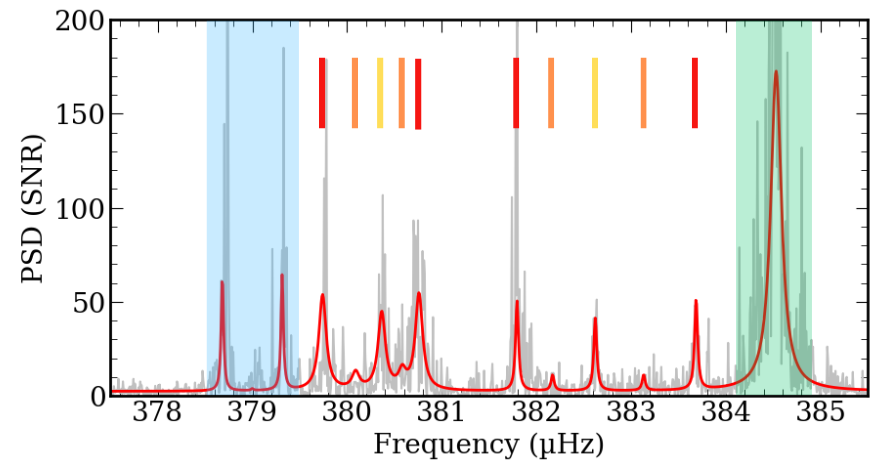}
        \textbf{(b)}
    \end{minipage}

    \begin{minipage}{\linewidth}
        \centering
        \includegraphics[width=\linewidth]{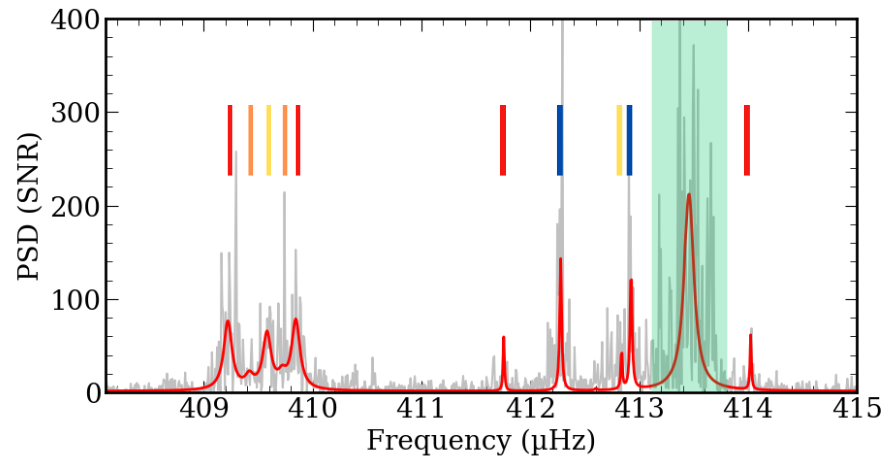}
        \textbf{(c)}
    \end{minipage}
    
    \caption{Asymptotic fit of KIC 7341231. In both panels, the \textit{Kepler} data is in grey and our asymptotic fit in red. Radial, $\ell=1$, and $\ell=2$ modes are highlighted by the shading in green, blue, and yellow, respectively. (a) Two orders showing mixed $\ell=2$ modes with asymmetric splittings; (b) Zoom near 411 µHz on split $\ell=2$ modes; $m=0$, $\pm1$, and $\pm2$ components are marked in yellow, orange, and red. (c) Zoom near 381 µHz on split $\ell=2$ modes; model in thick red, data in grey; $m=0$, $\pm1$, and $\pm2$ components of the $\ell=2$ mode are marked in yellow, orange, and red and $m=\pm1$ components of the $\ell=1$ mode are marked in dark blue. }
    \label{fig:Dh17_combined}
\end{figure}

\begin{figure}[ht]
    \centering

    \begin{minipage}{\linewidth}
        \centering
         \includegraphics[width=\linewidth]{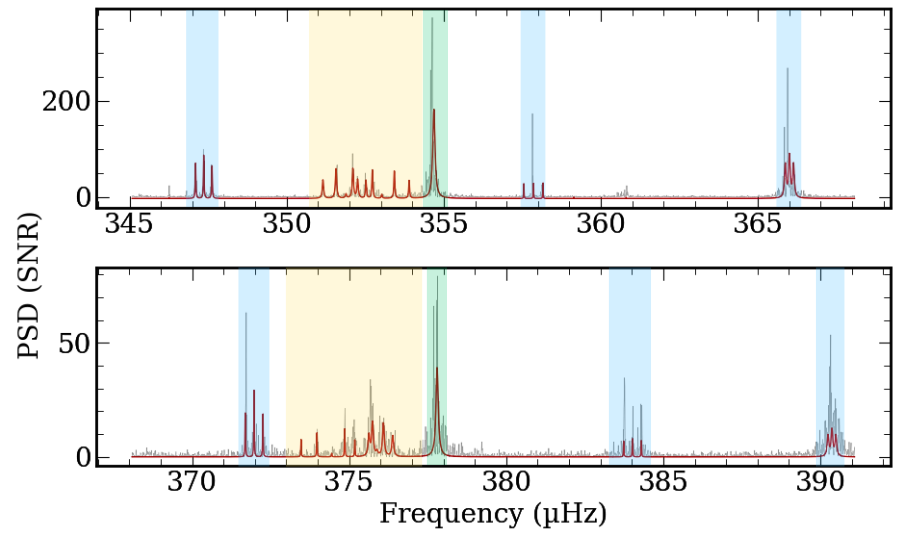}
        \textbf{(a)}
    \end{minipage}
    
    \vspace{0.5cm} 
    
    \begin{minipage}{\linewidth}
        \centering
        \includegraphics[width=\linewidth]{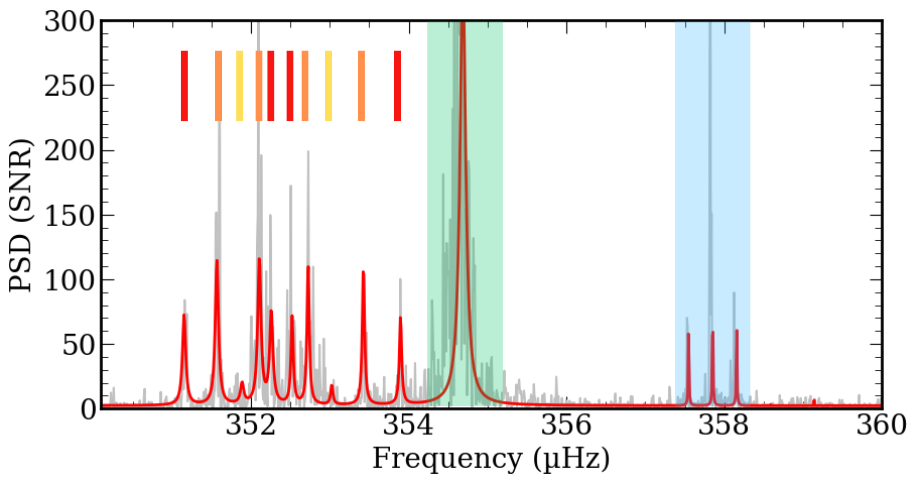}
        \textbf{(b)}
    \end{minipage}
    
    \caption{Asymptotic fit of KIC 8179973. In both panels, the \textit{Kepler} data is in grey and our asymptotic fit in red. Radial, $\ell=1$, and $\ell=2$ modes are highlighted by the shading in green, blue, and yellow, respectively. (a) Two orders showing mixed $\ell=2$ modes with asymmetric splittings; (b) Zoom near 352 µHz on split near-degenerate $\ell=2$ modes. The quadrupolar mixed $m=0$, $\pm1$, and $\pm2$ components are marked in yellow, orange, and red, respectively. The full fit can be found in Fig.~\ref{fig:echelle_Alice}}
    \label{fig:Alice_combined}
\end{figure}

Figures~\ref {fig:Dh17_combined} and \ref{fig:Alice_combined} show the result of the fitting procedure described in Section~\ref{sec:priors} obtained for the two stars. The results are reported in Table \ref{tab:res_quadfit} along with the $1\sigma$ uncertainties extracted from the Bayesian analysis (the cornerplots of the final fits can be found in App. \ref{app:corner}). When the posterior was asymmetrical, we chose to report the highest value of uncertainty as the error-bar.

\begin{table}[ht]
\caption{Best fit parameters for KIC 7341231 and KIC 8179973 with $1\sigma$ uncertainties.}
\begin{tabular}{l|ll}
\hline\hline
\diagbox{Param.}{Star} & KIC 7341231       & KIC 8179973       \\ \hline
$\numax$ (µHz)                               & $408\pm 15$       & $347\pm12$        \\
$\dnu$ (µHz)                                 & $29.00\pm0.01$    & $23.00\pm0.01$    \\
$\varepsilon$                                & $0.256\pm0.004$   & $0.425\pm0.006$   \\
$\delta\nu_{0,1}$ (µHz)                      & $-0.118\pm0.02$   & $-0.346\pm0.016$  \\
$\delta\nu_{0,2}$ (µHz)                      & $3.413\pm0.021$   & $2.182\pm0.025$   \\
$\beta_{0,2}$                                & $0.009\pm0.001$   & $0.011\pm0.001$   \\
$\Delta\Pi_1$ (s)                            & $111.180\pm0.025$ & $97.011\pm 0.008$ \\
$q_1$                                        & $0.379\pm0.001$   & $0.197\pm0.001$   \\
$\varepsilon_{g,1}$                          & $0.310\pm0.006$   & $0.279\pm0.003$   \\
$\Delta\Pi_2$ (s)                            & $64.422\pm0.007$  & $56.015\pm0.023$  \\
$q_2$                                        & $0.040\pm0.001$   & $0.020\pm0.001$   \\
$\varepsilon_{g,2}$                          & $0.138\pm0.004$   & $0.135\pm0.02$   \\
${\Omega_\mathrm{core}}/{2\pi}$ (nHz)        & $781\pm 4$        & $672\pm8$         \\
${\Omega_\mathrm{env}}/{2\pi}$ (nHz)         & $53\pm6$          & $49\pm8$          \\
$\theta_\mathrm{inc}$ (°)                    & $85$ (fixed)      & $51\pm3$          \\
$\alpha$                                     & $0.009\pm0.001$   & $0.006\pm0.001$   \\
$\beta_{0,1}$                                & $0.008\pm0.001$   & $0.004\pm0.001$   \\ \hline
\end{tabular}

\label{tab:res_quadfit}
\end{table}

\subsubsection{KIC 7341231}

In Figure~\ref{fig:Dh17_combined}, we show our global fit of the mixed modes of KIC 7341231, and highlight in panels (b) and (c) the identification of $\ell=2$ modes on two consecutive radial orders. We vet the asymptotic formula and fitting framework by comparing the results with the work of \cite{2017A&A...605A..75D}, which made use of complete evolutionary models of the star to analyse the radial order represented in panel (b). We show that our fit is consistent with all visible $\ell=2$ modes in the PSD, as first identified in Fig.1 of \cite{2017A&A...605A..75D} in the neighbourhood of $380$ µHz. \\ 

Besides the specific radial order analysed by \cite{2017A&A...605A..75D} in the vicinity of $380$ µHz, the asymptotic fit allows systematic identification of $\ell=2$ modes across the four surrounding radial orders (see Fig.~\ref{fig:ech_all_dh17}), which is a first for this star. In addition, we are able to properly identify multiplet components even when modes overlap between different angular degrees $\ell$. For instance, in Fig. \ref{fig:Dh17_combined} (a) and (c), around 413 µHz, three modes of degree $\ell=$0,1,2 are superimposed, and each peak has been successfully identified individually by reconstruction of the asymptotic mode pattern. \\

In Table~\ref{tab:rot}, we show that our estimate of core and envelope rotation rates are in agreement with model-based measurements from \cite{2017A&A...605A..75D}, and consistent across radial orders. We obtain uncertainties of about 3 times smaller for core rotation rate, and 2 times smaller for the envelope rotation rate: this is due to the fact that our fit is global \citep[as also performed for magnetic effects in][]{2024MNRAS.534.1060H}. Indeed, uncertainties in mixed-mode parameters are usually estimated through non-parametric methods that do not exploit the full spectral information and instead consider errors on individual frequencies rather than the coherence of the global mixed-mode pattern \citep[\textit{e.g.}][]{2018A&A...616A..24G,2023ApJ...954..152K}. The global Bayesian fit is also delivering better precision on asymptotic parameters in Table~\ref{tab:res_quadfit}. For this reason, we advocate for a broad use of these global fitting methods.
\\

\begin{table}[ht!]
\centering
\caption{Comparison of the rotation rates found in this work and  \citet[][Dh17 in the table]{2017A&A...605A..75D}.}
\label{tab:rot}
\begin{tabular}{llll}
\hline\hline
Parameter &      & This work & Dh17 \\ \hline
${\Omega_\mathrm{core}}/{2\pi}$ &(nHz)   & $781\pm4$   & $771\pm13$                   \\
${\Omega_\mathrm{env}}/{2\pi}$ &(nHz)   & $53\pm6$        & $45\pm12$               \\\hline 
\end{tabular}
\end{table}
~
In order to assess the gain from adding constraints from $\ell=2$ modes to the fitting procedure, we estimate the Bayesian Information Criterion \citep[BIC,][]{10.1214/aos/1176344136} comparing the fit performed with $\ell=2$ mixed modes and $\ell=2$ p modes only. For that, we fit the data again with split mixed dipole modes and pressure $\ell=2$ modes as usually done in the community and compare it against the full fit of the present work. The inclusion of mixed $\ell=2$ modes in the model leads to a BIC that is 3,726 points lower than when using only mixed $\ell=1$ modes and pressure $\ell=2$ modes on the target KIC 7341231. Because differences in BIC approximate:
\begin{equation}
    \Delta\mathrm{BIC}=-2\ln(\mathrm{BF})\,,
\end{equation} 
where BF is the bayes factor; such a large $\Delta\mathrm{BIC}$ corresponds to overwhelmingly strong statistical evidence in favour of the extended model. We do not report the absolute BIC values themselves, as they are not tied to any absolute scale and carry no stand-alone interpretation; only their differences have statistical meaning for model comparison. This substantial improvement in BIC therefore indicates very strong support for including the mixed $\ell=2$ modes, suggesting that the extended model provides a significantly better representation of the observed oscillation data. As a result, the physical inferences drawn from this model (such as internal rotation rates) can be considered significantly more robust and better constrained than those inferred from mixed $\ell=1$ modes only.\\
\subsubsection{KIC 8179973}

We then perform the first fit of the radial and $\ell=1,2$ mixed-mode pattern for KIC 8179973. This star presents a particularly asymmetric pattern in the $\ell=2$ regions, as the multiplets are very close to each other (see panels (a) and (b) of Fig.~\ref{fig:Alice_combined}). The asymptotic prescription (Eq.~\ref{off-diag-factor}) yields a very good interpretation of the PSD, with all visible modes being accurately represented (see also Fig.~\ref{fig:echelle_Alice}). The values of $\Delta\Pi_1,\varepsilon_{g,1}$ and $q_1$ are also in the range of expected values from scaling relations and measurements of \citep{2017A&A...600A...1M,2023ApJ...954..152K}. The small number of visible mixed $\ell=2$ modes is expected from the small mixed mode density parameter \citep{2018A&A...616A..24G} $\mathcal{N}_2={\dnu}/({\Delta\Pi_2\numax^2})\approx 3.4$ and the fact that the coupling factor $q_2$ is very small.
Furthermore, for this star, we have used a much less informative prior for $q_2$ as in the fit of KIC 7341231, which shows the robustness of our fitting method. In addition, $\Delta\Pi_2$ and $\Delta\Pi_1$ are compatible within $2\sigma$ following Eq.~(\ref{eq:dpi}). From the global fit, we obtain $\Omega_{\rm core}=671\pm8$ µHz and $\Omega_{\rm env}=49\pm8$~µHz, which are within the expected range for a young red giant star \citep{2018A&A...616A..24G}. Like for KIC 7341231, the uncertainties we obtain on KIC 8179973 are significantly smaller than those typically reported in similar studies (see, \textit{e.g.}, \citealt[][]{2018A&A...616A..24G,2014A&A...564A..27D}). This improvement highlights the strength of our global fitting procedure, which (i) minimises the uncertainties by simultaneously exploiting information from multiple modes, and (ii) naturally mitigates the ambiguity between noise and genuine power excesses thanks to the intrinsic pattern constraints of the analysis. Compared to a fit that includes only the pure $\ell=2$ pressure modes, the BIC derived from the full asymptotic framework is lower by 3,185 points. This provides strong statistical evidence that the prescription including mixed quadrupole modes and incorporating near-degeneracy effects, offers a superior inference of the underlying physical parameters. This measurement of near-degeneracy effects in a second early red giant paves the way for a fully automated fitting pipeline based on asymptotic mixed-mode parameters across $\ell$.

\section{Conclusions}
\label{sec:ccl}

This work is the first to infer the complete set of asymptotic parameters of $\ell=2$ modes and the first fit of $\ell=2$ mode near-degeneracy splittings not relying on a numerical model of the structure of the star. In particular, it leads to the first inferences of $q_2$, $\Delta\Pi_2,$ and $\varepsilon_{{\rm g},2}$ and, with custom ways to constrain the coupling factor $q_2$, paves the way for an automation of the fit and identification of $\ell=2$ mixed modes.\\

Our results for $\Delta\Pi_1$ and $\Delta\Pi_2$ in KIC 8179973 are consistent within $2\sigma$. However, in KIC 7341231, the values agree within a few tenths of seconds with non-overlapping errors, suggesting either a small deviation from the asymptotic theory or the influence of surface effects that were not fully accounted for in this work. Indeed, because the fit of asymptotic mixed modes is very reliant on the position of the pure p-mode, a difference in the surface effects affecting the $\ell=1$ and $\ell=2$ notional p modes would suffice to explain a small difference in $\Delta\Pi$.
In addition, our asymptotic formulation, including the $\ell=2$ modes, has enabled us to put stronger constraints on the rotation rates of our targets. Indeed, in KIC 7341231 we measured an envelope (resp. core) rotation rate of $57\pm11$~nHz (resp. $777\pm10$~nHz) with only $\ell=1$ modes, while with the addition of $\ell=2$ modes combined with a global fitting framework, we obtain an envelope (resp. core) rotation of $53\pm6$~nHz (resp. $781\pm4$ nHz). Those values are compatible with each other, and we also observe that the inclusion of the mixed $\ell=2$ modes makes the uncertainty drop by a factor of $\approx$2 ($2.5$ for the core rotation and $1.8$ for envelope rotation). This is mostly due to the inclusion of more modes in the fitting procedure. There might also be a different contribution from the sensitivity kernels of the $\ell=2$ modes. Specifically, their shallower turning points offer distinct constraints from those of $\ell=1$ modes in the p-mode cavity, suggesting a potential for improved radial resolution in future inversions. This strengthening of the constraints on rotation rates is further supported by the fact that including the mixed $\ell=2$ modes made the BIC of the fit of KIC~7341231 and KIC~8179973 drop by more than 3,000 points, which is a strong statistical evidence that the use of the new formulation allowing to fit $\ell=2$ mixed modes leads to better physical inference. 
In addition, as we have shown in Fig. \ref{fig:Dh17_combined}, our fitting framework can distinguish between entwined split modes of different $\ell$, which shows the robustness of the fitting methodology.
\\

The asymptotic framework is also expected to simplify future studies of magnetic signatures and yield new results using not only the $\ell=1$ but also the $\ell=2$ modes to characterise magnetic fields in the RGB. We expect that the measurement of $q_2$ will help constrain the mode suppression phenomenon \citep[\textit{e.g.}][]{2015Sci...350..423F,2016Natur.529..364S,2017A&A...598A..62M} for $\ell=2$ modes \citep{2016PASA...33...11S} as the loss in power of weakened non radial modes modes is expected to be linked to the transmission coefficient from the p mode cavity to the g mode cavity \citep{2015Sci...350..423F,2020MNRAS.493.5726L,2025A&A...696A.134M}. Low and radial-order-dependent values of $\Delta\Pi_2$ should be a hint of the presence of an internal magnetic field, as is it done with $\Delta\Pi_1$ \citep{2020MNRAS.496.3829L, 2022A&A...667A..68B,2023A&A...670L..16D}. In a follow-up study (Liagre et al., in prep.), we will employ this new formulation and framework together with first-order magnetic effects \citep[\textit{e.g.}][]{2020MNRAS.496..620G,2021A&A...650A..53B,2021A&A...647A.122M,2022A&A...667A..68B,2022Natur.610...43L,2023A&A...676L...9M,2024ApJ...970...42B,2024A&A...690A.217D,2023A&A...680A..26L,2024MNRAS.534.1060H,2024ApJ...970...42B} to disentangle rotation-induced near-degeneracy asymmetries from magnetic ones in $\ell=2$ modes. This is particularly important, as it should provide stronger constraints on the magnetic field topology, as shown by \citet{2024A&A...690A.217D}.\\

We anticipate that ongoing and upcoming space missions will continue to expand the sample of stars with good signal-to-noise ratios observed in this intermediate evolutionary phase. During its extended operations, the Transit Exoplanet Survey Satellite \citep[TESS,][]{2014SPIE.9143E..20R} mission improved its temporal resolution, reducing the full-frame image cadence first to 10 minutes and later to 3 minutes. The TESS data are currently being calibrated for seismology \citep[][García et al., in prep.]{2024tkas.confE.123G}, which will enable large-scale and homogeneous seismic analyses. Moreover, the upcoming PLAnetary Transits and Oscillations of stars (PLATO; \citealt{2014ExA....38..249R}) mission is expected to enable the detection of a large number of solar-like oscillators in the subgiant and early giant phases from the calibration sample, consistent with recent yield predictions \citep{2024A&A...683A..78G}. Using this global asymptotic description along with the Bayesian fitting framework, we expect not only to characterize with an increased precision the rotation profile of stars exhibiting mixed $\ell=2$ modes but also of those more evolved stars, exhibiting near-degenerate $\ell=1$ modes \citep[using the full formalism or for instance relying on Eq.~(\ref{rotation_system}), see also][]{2025arXiv250926319A}.

\begin{acknowledgements}
We thank the referee for their careful and constructive report, which has substantially enhanced both the quality and clarity of the manuscript. L. Bugnet and L. Einramhof gratefully acknowledge support from the European Research Council (ERC) under the Horizon Europe programme (Calcifer; Starting Grant agreement N$^\circ$101165631). While partially funded by the European Union, views and opinions expressed
are, however, those of the authors only and do not necessarily reflect those of the European Union or the European Research Council. Neither the European Union nor the granting authority can be held responsible for them. The authors acknowledge the great support and feedback provided during the redaction of this article by Pr. Rafael Garc\'ia and Pr. Savita Mathur. We would also like to thank Dr. Emily Hatt for her insights on uncertainty estimates. The authors also thank the members of the Asteroseismology and Stellar Dynamics group of the Institute of Science and Technology Austria (ISTA) for very useful discussions: L. Barrault, S.B. Das, K. Smith.

This paper includes data collected by the \emph{Kepler} mission and obtained from the MAST data archive at the Space Telescope Science Institute (STScI). Funding for the \emph{Kepler} mission is provided by the NASA Science Mission Directorate. STScI is operated by the Association of Universities for Research in Astronomy, Inc., under NASA contract NAS 5–26555.
\\
\textit{Software:} AstroPy \citep{astropy:2013,astropy:2018}, Matplotlib \citep{Hunter:2007}, NumPy \citep{harris2020array}, SciPy \citep{2020SciPy-NMeth}, emcee \citep{2013PASP..125..306F}, celerite \citep{celerite}, slepc4py \citep{DALCIN20111124,10.1145/1089014.1089019}, KADACS \citep{2011MNRAS.414L...6G}, sloscillations \citep{2019MNRAS.488..572K,2023ApJ...954..152K}
\end{acknowledgements}

\bibliographystyle{aa}
\bibliography{BIBLIO_ND}
\begin{appendix}
\section{Perturbative treatment of the near degeneracy effects}\label{pert_treat}
Although not used for our study, we provide in this appendix an explicit approximation to the solution of the eigenvalue problem, as it could be used as a simpler and faster implementation of the near degeneracy effects when they are not strong.
In particular, we show how we can use perturbative methods to approximate the eigenvalues of the quadratic eigenvalue problem (Eq.~\ref{eq:mat_eigval}). This approach provides closed-form solutions that are valid in the regime of weak near-degeneracy \textit{i.e.} when the off-diagonal entries of $R$ are small compared to the diagonal one but not negligible (typically one order of magnitude less). In addition, we demonstrate that it offers a new way to determine both the core and envelope rotation rates from a single near-degeneracy–asymmetric splitting.\\

The quadratic eigenvalue problem is equivalent to solving the set of coupled equations:
\begin{equation}
    -\omega^2a_i+\omega\sum_j R_{ij}a_j+\omega_{0,i}^2a_i=0\,, \forall i.
\end{equation}
Since at first order the splittings are symmetrical and the off diagonal terms decrease rapidly, we expect $\sum R_{ij}a_j\sim R_{ii}a_i$ and because $\lvert R_{ii}\rvert\ll\omega_{0,i}$, at the zeroth order $\omega=\omega_{0,i}$ (where we chose the positive branch of solutions). Hence, injecting the zeroth-order solution into each equation of the system yields that it can be approximated by:
\begin{equation}
    -\omega^2a_i+\omega_{0,i}\sum_j R_{ij}a_j+\omega_{0,i}^2a_i=0\,,\forall i.
\end{equation}
This can be expressed as a matrix equation such as:
\begin{equation}
    A\Vec{a}=\omega^2\Vec{a}\,,
\end{equation}
with:
\begin{equation}
    A_{ij} = \omega_{0,i}^2\,\delta_{i,j}
+ \omega_{0,i}R_{ij}\,.
\end{equation}
Then, using quantum perturbation theory, we find an explicit approximate solution to this eigenvalue problem, yielding the following expression for the frequencies:
\begin{equation}
\begin{split}
    \nu_{i,m}^2 \approx {} & \nu_{i,0}^2 \\
    & + 2m\nu_{i,0}\left(
        \zeta_i\left(1-\frac{1}{L^2}\right)\frac{\Omega_\mathrm{core}}{2\pi}
        + (1-\zeta_i)\frac{\Omega_\mathrm{env}}{2\pi}
    \right) \\
    & + \left(
        \frac{\Omega_\mathrm{core}}{2\pi}
        - \frac{L^2}{L^2-1}\frac{\Omega_\mathrm{env}}{2\pi}\right)^2
    \sum_{i\neq j}
    \frac{4m^2\nu_{i,0}^2\gamma_{c,ij}^2}{\nu_{i,0}^2-\nu_{j,0}^2}
\end{split}
\label{pert_eigen}
\end{equation}
This shows that the asymmetry in the splitting due to the near degeneracy is a second order effect. This formulation allows us to define an asymmetry parameter for one splitting:
\begin{equation}
    \label{asym_param} 
    \begin{split}\delta_\mathrm{asym}&=\sqrt{\abs{\frac{\nu_{i,-m}^2+\nu_{i,m}^2-2\nu_{i,0}^2}{8m^2\nu_{i,0}^2}}}\\&=\sqrt{\sum_{i\neq j}
    \frac{\gamma_{c,ij}^2}{\abs{\nu_{i,0}^2-\nu_{j,0}^2}}}\left(
        \frac{\Omega_\mathrm{core}}{2\pi}
        - \frac{L^2}{L^2-1}\frac{\Omega_\mathrm{env}}{2\pi}\right)\\
        &=\alpha\left(
        \frac{\Omega_\mathrm{core}}{2\pi}
        - \frac{L^2}{L^2-1}\frac{\Omega_\mathrm{env}}{2\pi}\right)\,.
    \end{split}
\end{equation}
We can also recover the symmetric splitting that would occur without near-degeneracy effects:
\begin{equation}
    \label{sym_split} 
    \begin{split}
    \delta\nu_\mathrm{sym}&=\frac{\nu_{i,m}^2-\nu_{i,-m}^2}{4\abs{m}\nu_{i,0}}\\&=\left(1-\frac{1}{L^2}\right)\zeta_i\frac{\Omega_\mathrm{core}}{2\pi}+ (1-\zeta_i)\frac{\Omega_\mathrm{env}}{2\pi}\,.
    \end{split}
\end{equation}
If we further define $\beta=\frac{L^2}{L^2-1}\alpha$, $\eta=\frac{L^2-1}{L^2}\zeta_i$, $\lambda=(1-\zeta_i)$, the 2 parameters $\delta_\mathrm{asym}$ and $\delta\nu_\mathrm{sym}$ can be linearly combined to obtain the core and envelope rotational rates from only one splitting. The appropriate linear combinations are the following:
\begin{equation}
\begin{cases}
    \dfrac{\Omega_\mathrm{core}}{2\pi}
    = \dfrac{1}{\alpha\lambda+\beta\eta}\left(\lambda\delta_\mathrm{asym}+\beta\delta\nu_\mathrm{sym}\right) \\
    \dfrac{\Omega_\mathrm{env}}{2\pi}
    = \dfrac{1}{\alpha\lambda+\beta\eta}\left(-\eta\delta_\mathrm{asym}+\alpha\delta\nu_\mathrm{sym}\right)\,.
\end{cases}
\label{rotation_system}
\end{equation}
This shows that in the case of a splitting with near-degeneracy effects, both the envelope and core rotation rates are encoded simultaneously in the morphology of the splitting. Whereas, without near-degeneracy effects as usually studied in $\ell=1$ modes, we can only access a linear combination of both rotation rates and two multiplet  (ideally one p- and one g- dominated) are needed to constraint both the core and the envelope rotations.
\section{Vetting the asymptotic formulation}
\begin{figure}[h]
    \centering
    \includegraphics[width=\linewidth]{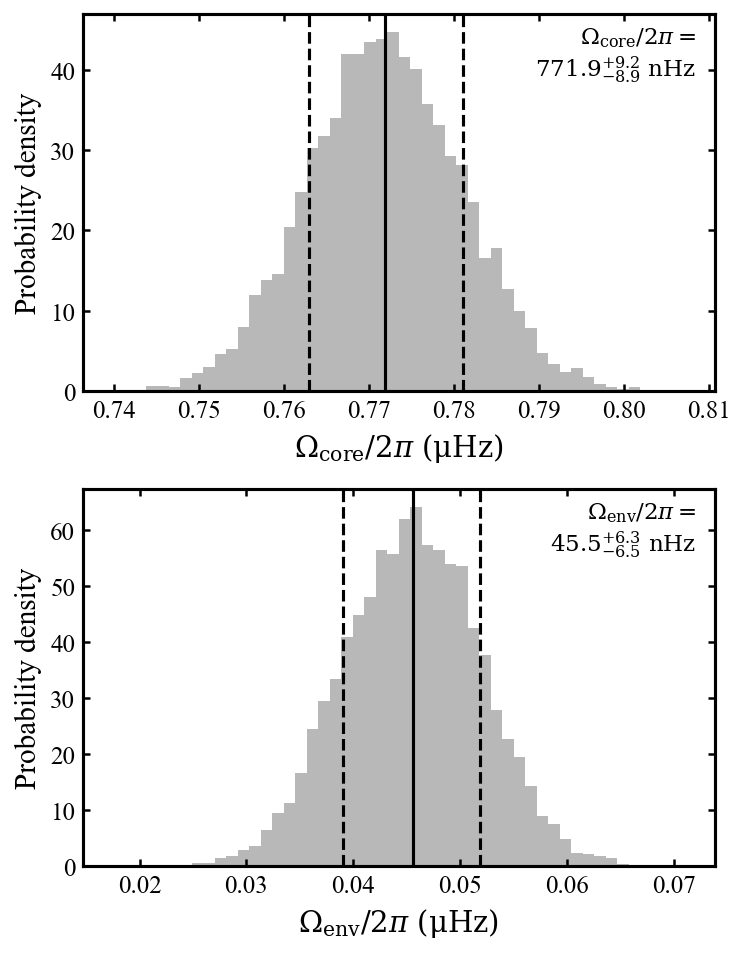}
    \caption{Histograms showing the result of the Monte-Carlo analysis with the asymptotic formulation}
    \label{fig:reproduction_Dh}
\end{figure}
To propagate the observational uncertainties of the mixed-mode parameters into the inferred rotational components, we performed a Monte-Carlo analysis with \(N = 10{,}000\) realizations. For each realization, we randomly drew the parameters \((\nu_a, \nu_b, \Delta\Pi_2, \zeta_a, \zeta_b)\) from independent Gaussian distributions centered on the value reported in \cite{2017A&A...605A..75D} and with their respective uncertainties. For every random draw, we then minimized the \(\chi^2\) difference between the observed multiplet frequencies and the model frequencies computed following Sec. \ref{sec:dh_test}, treating \(\Omega_\mathrm{core}\) and \(\Omega_{\rm env}\) as free parameters. The optimization was carried out using the L-BFGS-B algorithm, with the bounds \(\nu_c \in [0,1]~\mu{\rm Hz}\) and \(\nu_{\rm env} \in [0,0.1]~\mu{\rm Hz}\). Only successful minimizations were retained in the final sample. The resulting posterior distributions of \(\nu_c\) and \(\nu_{\rm env}\) were summarized by their median values and their 16th–84th percentile ranges, which we report as our final uncertainties.

\section{Bayesian priors}
\label{app:priors}

\subsection{Priors on pressure mode and mixed $\ell=1$ modes parameters}\label{prior_pl1}

$\numax$ is fixed to the value given by PyA2Z while we assign flat priors within $\pm5\%$ of the initial guess for $\dnu$ and $\Delta\Pi_1$ to account for the uncertainties of the PyA2Z pipeline and we choose not to constrain $\varepsilon$ as its value is highly sensitive to any change in $\dnu$.
We assign the priors on $\delta\nu_{0,\ell}$ around the values predicted by scaling relations provided by \citet{2018A&A...618A.109M,2012ApJ...757..190C}.
We assign to $q_1$ a flat prior between $0.15$ and $0.4$ based on the results of previous studies such as \cite{2017A&A...600A...1M, 2023ApJ...954..152K}.
For $\varepsilon_{{\rm g},1}$, we chose a very uninformative flat prior between $0$ and $1$. In the same fashion we chose very uninformative flat priors for $\beta_{0,\ell}$ and $\alpha$ between $-0.1$ and $0.1$. This range seemed appropriate as the $\alpha$ and $\beta_{0,\ell}$ parameters found by \citet{2020A&A...642A.226A} and \citet{2017ApJ...835..172L} in the same evolutionary stage were under $10^{-2}$.
We assigned flat priors for rotation rates, with values ranging from $0$ to $1,000$ nHz for the core and $0$ to $100$ nHz for the envelope. These core rotation prior was derived from the work of \citet{2017A&A...605A..75D} for KIC 7341231 and informed by a visual inspection of the splittings observed in KIC 8179973. For the envelope rotation, we again utilized the \citet{2017A&A...605A..75D} article for KIC 7341231, and assumed a core envelope contrast of at least 10 in KIC 8179973.
All of those priors are summarized in table \ref{tab:priors}.

\subsection{Priors on KIC 7341231}
\label{Prior_7341231}
A global analysis using PyA2Z gives $\dnu$, $\numax$, and a guess value of $\Delta\Pi_1$ using the stretched period method. The priors on those parameters were then chosen according to Sec \ref{prior_pl1}. Refined values of $\Delta\Pi_1$ and $\dnu$ are obtained by fitting the mixed $\ell=1$ modes according to the method described in Sec. \ref{Bayesfit}.Because the angle of inclination is high, we had access to the $m=0$ components of each quintuplet. Hence, we decided to use the method described in \ref{q_from_modes} to constrain $q_2$. To use this method, we needed to access the following parameters and their associated uncertainties: $\Delta\Pi_2$, $\dnu$, $\nu_a,\nu_b$ and the position of the $\ell=2$ p mode that is given by $\nu_0-\delta\nu_{0,2}$. Where $\nu_a$ and $\nu_b$ are the positions of the $m=0$ components of each multiplet and $\nu_0$ is the frequency of the closest radial mode.
$\Delta\Pi_2$ and its associated uncertainty was then obtained following Sec.\ref{prior_l2}.
In order to obtain $\delta\nu_{0,2}$, we fitted a bimodal Gaussian model to the averaged collapse of the echelle diagram. We obtained ${\delta\nu_{0,2}}/{\dnu}=0.120\pm0.002$.
We finally peakbagged $\nu_a,\nu_b$ and $\nu_0$ and obtained $\nu_a=380.382\pm0.023$ µHz,$\nu_b=382.627\pm0.01$ µHz, and $\nu_0=384.514\pm0.011$ µHz. Using those values and their uncertainties, we can use a Monte-Carlo method to sample the distribution of $q_2$ according to equation (\ref{q_ana}). We obtained $q_2=0.041\pm0.003$. We decided to use a uniform prior on $q_2$ within 3 sigma of the obtained value.\\
Furthermore, since we are only interested in reproducing the frequencies of the star's modes and an angle of inclination was already fitted by \cite{2012ApJ...756...19D}, we adopt a fixed angle of inclination of $85$°, in order to limit the number of free parameters.

\begin{table}[ht]
\caption{Flat prior ranges for the global fit of KIC 7341231 and KIC 8179973.}
\begin{tabular}{l|ll}
\hline\hline
\diagbox{Prior}{Star} & KIC 7341231       & KIC 8179973       \\ \hline
$\numax$ (µHz)                              & 408 (fixed)        & 347 (fixed)        \\
$\dnu$ (µHz)                                & $29.1\pm 5\%$      & $23\pm 5\%$        \\
$\varepsilon$                               & 0 < 1              & 0 < 1              \\
$\delta\nu_{0,1}$ (µHz)                     & -1 < 1             & -1 < 1             \\ 
$\delta\nu_{0,2}$ (µHz)                     & 2 < 3              & 2 < 3              \\
$\beta_{0,2}$                               & -0.1 < 0.1         & -0.1 < 0.1         \\
$\Delta\Pi_1$ (s)                           & 104.5 < 115.5      & 92 < 102           \\
$q_1$                                       & 0.15 < 0.4         & 0.15 < 0.4         \\
$\varepsilon_{{\rm g},1}$                         & 0 < 1              & 0 < 1              \\
$\Delta\Pi_2$ (s)                           & $\Delta\Pi_1/\sqrt{3} \pm 5\%$ & $\Delta\Pi_1/\sqrt{3} \pm 5\%$ \\
$q_2$                                       & 0.041 ± 0.009      & < 0.107            \\
$\varepsilon_{{\rm g},2}$                         & 0 < 1              & 0 < 1              \\
${\Omega_\mathrm{core}}/{2\pi}$ (nHz)       & <$1,000$              & <$1,000$              \\
${\Omega_\mathrm{env}}/{2\pi}$ (nHz)        & <100               & <100               \\
$\alpha$                                    & -0.1 < 0.1         & -0.1 < 0.1         \\
$\beta_{0,1}$                               & -0.1 < 0.1         & -0.1 < 0.1         \\ \hline
\end{tabular}

\label{tab:priors}
\end{table}

\subsection{Priors on KIC 8179973}
\label{Prior_8179973}
 A global analysis using PyA2Z gives $\dnu$, $\numax$, and a guess value of $\Delta\Pi_1$ using the stretched period method. The priors on those parameters were then chosen according to Sec \ref{prior_pl1}. Refined values of $\Delta\Pi_1$, $\dnu$ and a value of $q_1$ are obtained by fitting the mixed $\ell=1$ modes according to the method described in Sec. \ref{Bayesfit}. In KIC 8179973, the angle of inclination is such that the m=0 modes of the $\ell=2$ quintuplet are either absent or buried in the noise of the star. Hence, we cannot use the same method as the on used for KIC 7341231 to constrain $q_2$. However, we can still use simple analytical arguments to put a rough upper bound on $q_2$ as a function of $q_1$.\\
$q_1$ in itself is obtained by fitting the mixed $\ell=1$ modes according to the method described in Sec. \ref{Bayesfit}. We obtain $q_1=0.197\pm0.001$ which yields that $q_2<0.107$ using the arguments presented in Sec.\ref{up_lim_q2}.
We use a uniform prior for the orientation of axis of rotation of the star with respect to the line of sight. This translates to a prior $\propto\sin(\theta_\mathrm{inc})$ for $\theta_\mathrm{inc}$ between $0^\circ$ and $90^\circ$.

\subsection{Cornerplots and comparison data/model}\label{app:corner}

\begin{figure}[h!]
    \centering
    \includegraphics[width=\linewidth]{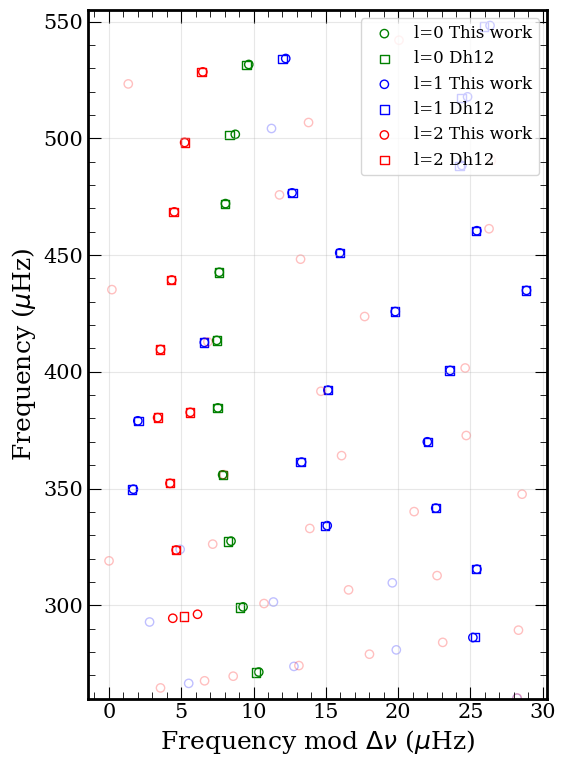}
    \caption{Echelle diagram comparing the oscillation frequencies derived from the Bayesian asymptotic model (circles, this work) with the peakbagged frequencies of KIC 7341231 (squares, \citealt{2012ApJ...756...19D,2017A&A...605A..75D}, referred to as Dh12 in the legend of the figure). The markers for the model frequencies are rendered semi-transparent when the separation from a peakbagged frequency exceeds 1~µHz.}
    \label{fig:Comparison_Dh12_17_model}
\end{figure}

We compared the frequencies obtained within our asymptotic Bayesian framework to the frequencies obtained by \citealt{2012ApJ...756...19D,2017A&A...605A..75D} using a peakbagging code. Both the asymptotic model and the peakbagged frequencies agree, particularly around the central frequency ($\numax$) as can be seen on figure \ref{fig:Comparison_Dh12_17_model}. This congruence is expected, as the asymptotic approximation becomes less precise further from $\nu_{\text{max}}$ due to increasing sensitivity to surface effects. Crucially, the physically informed nature of our method allows it to account for small excesses of power that are often indistinguishable from noise in standard peakbagging procedures.
\begin{figure*}
    \centering
    \includegraphics[width=0.61\linewidth]{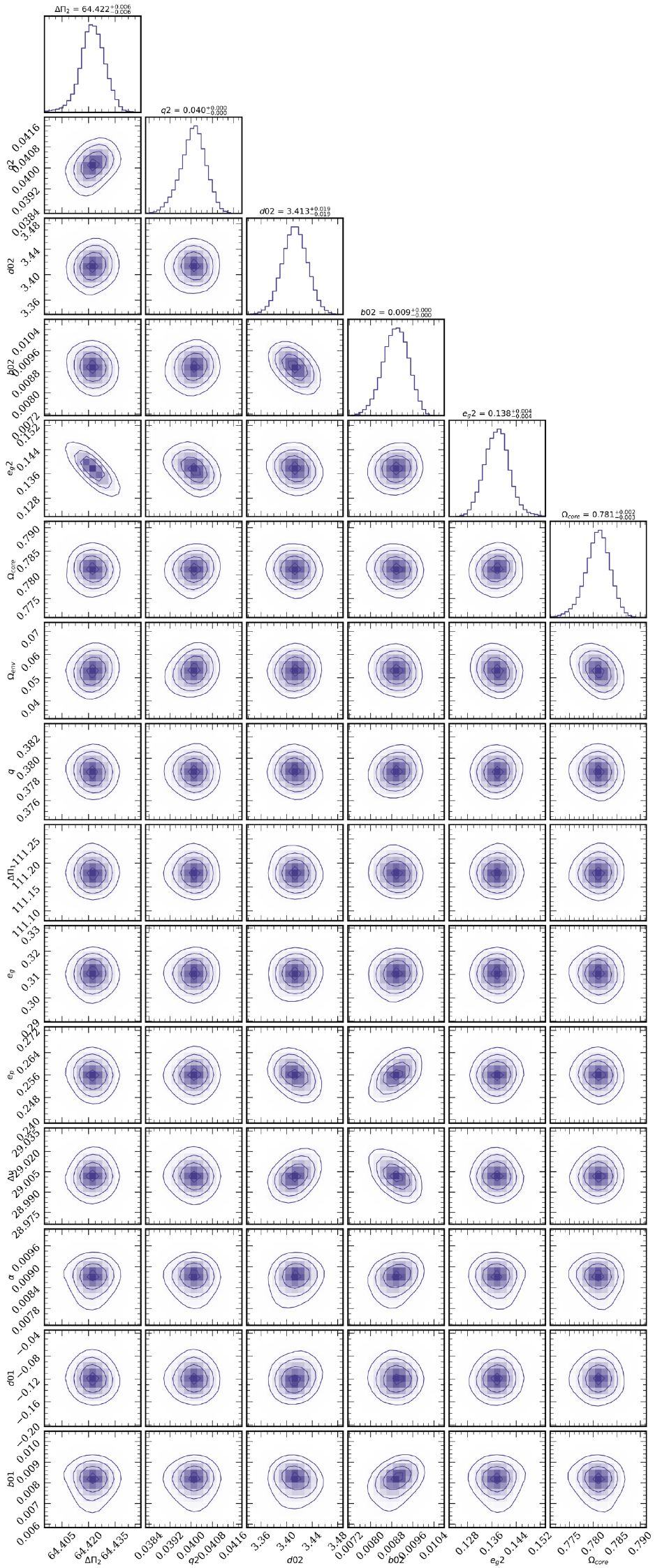}
    \caption{Cornerplot of the Bayesian fit of all the parameters for KIC 7341231 (first part).}
    \label{fig:corner_all_dh17_1}
\end{figure*}
\begin{figure*}
    \centering
    \includegraphics[width=\linewidth]{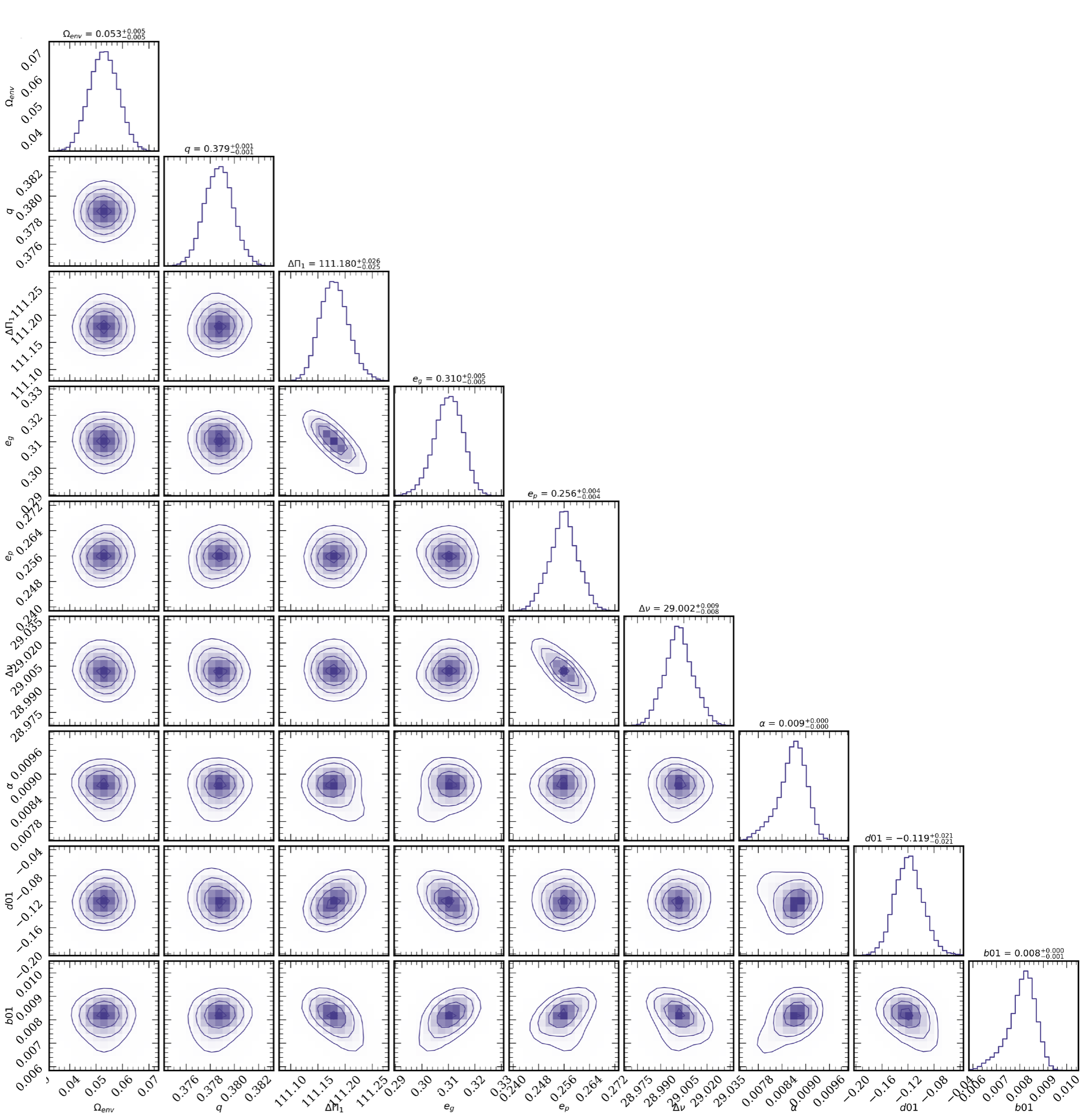}
    \caption{Cornerplot of the Bayesian fit of all the parameters for KIC 7341231 (second part).}
    \label{fig:corner_all_dh17_2}
\end{figure*}
\begin{figure*}
    \centering
    \includegraphics[width=0.9\linewidth]{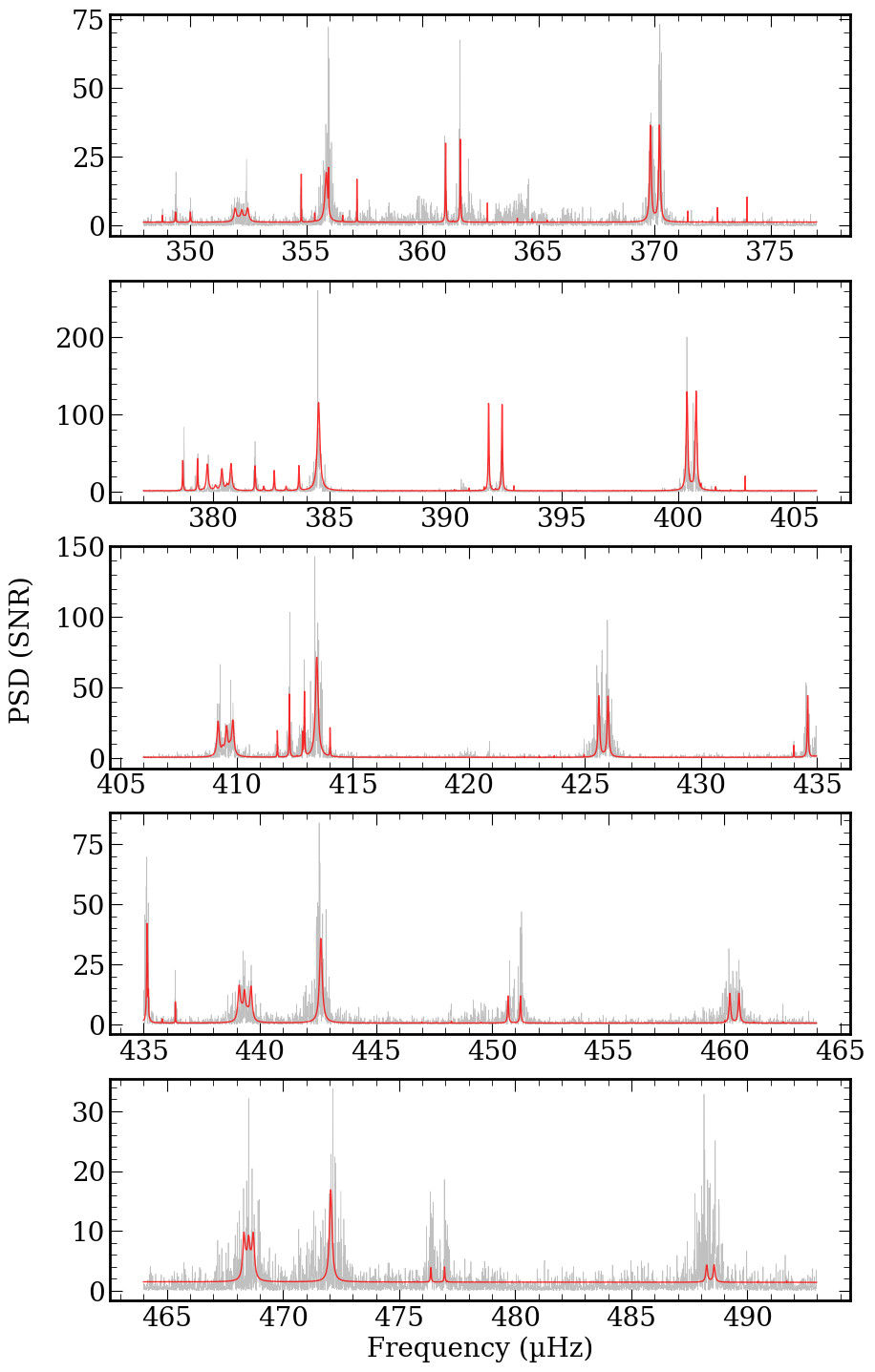}
    \caption{Echelle diagram showing the superimposed data and asymptotic model for the most prominent modes of KIC 7341231.}
    \label{fig:ech_all_dh17}
\end{figure*}
\begin{figure*}
    \centering
    \includegraphics[width=0.57\linewidth]{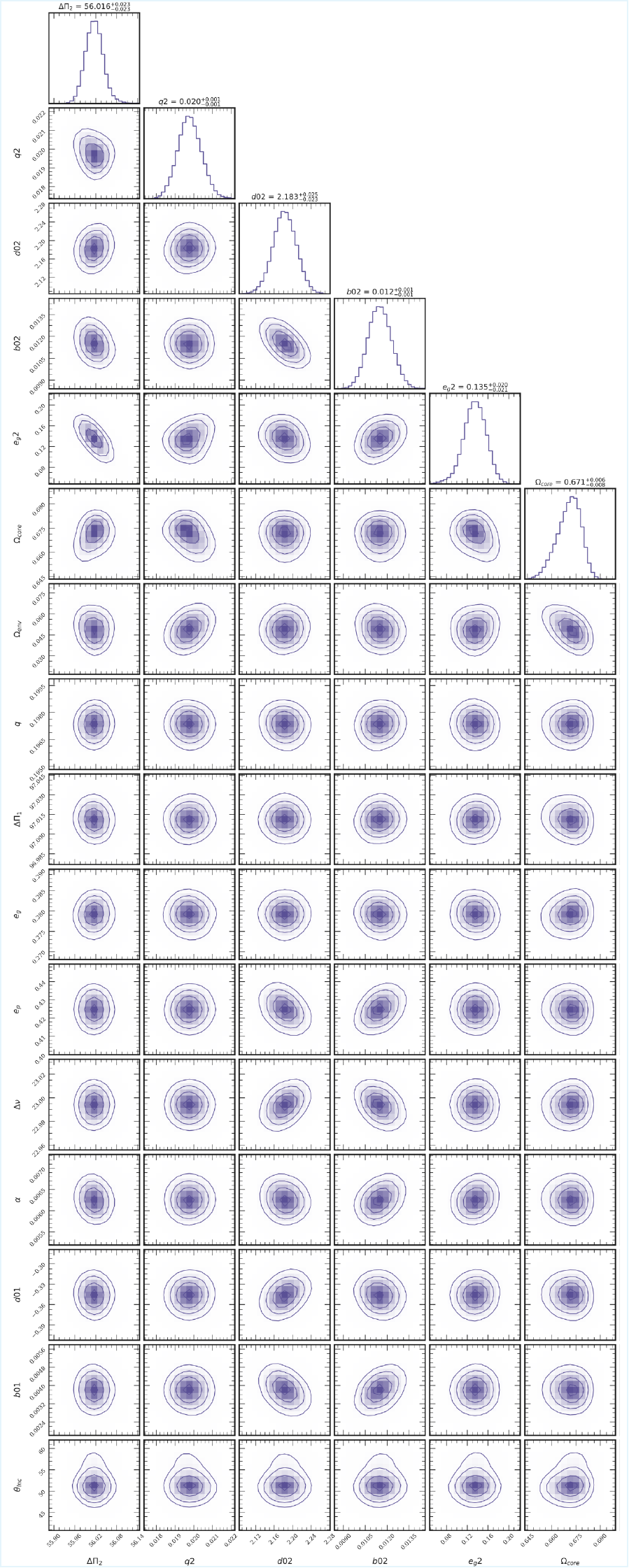}
    \caption{Cornerplot of the Bayesian fit of all the parameters for KIC 8179973 (first part).}
    \label{fig:corner_all_Alice_1}
\end{figure*}
\begin{figure*}
    \centering
    \includegraphics[width=\linewidth]{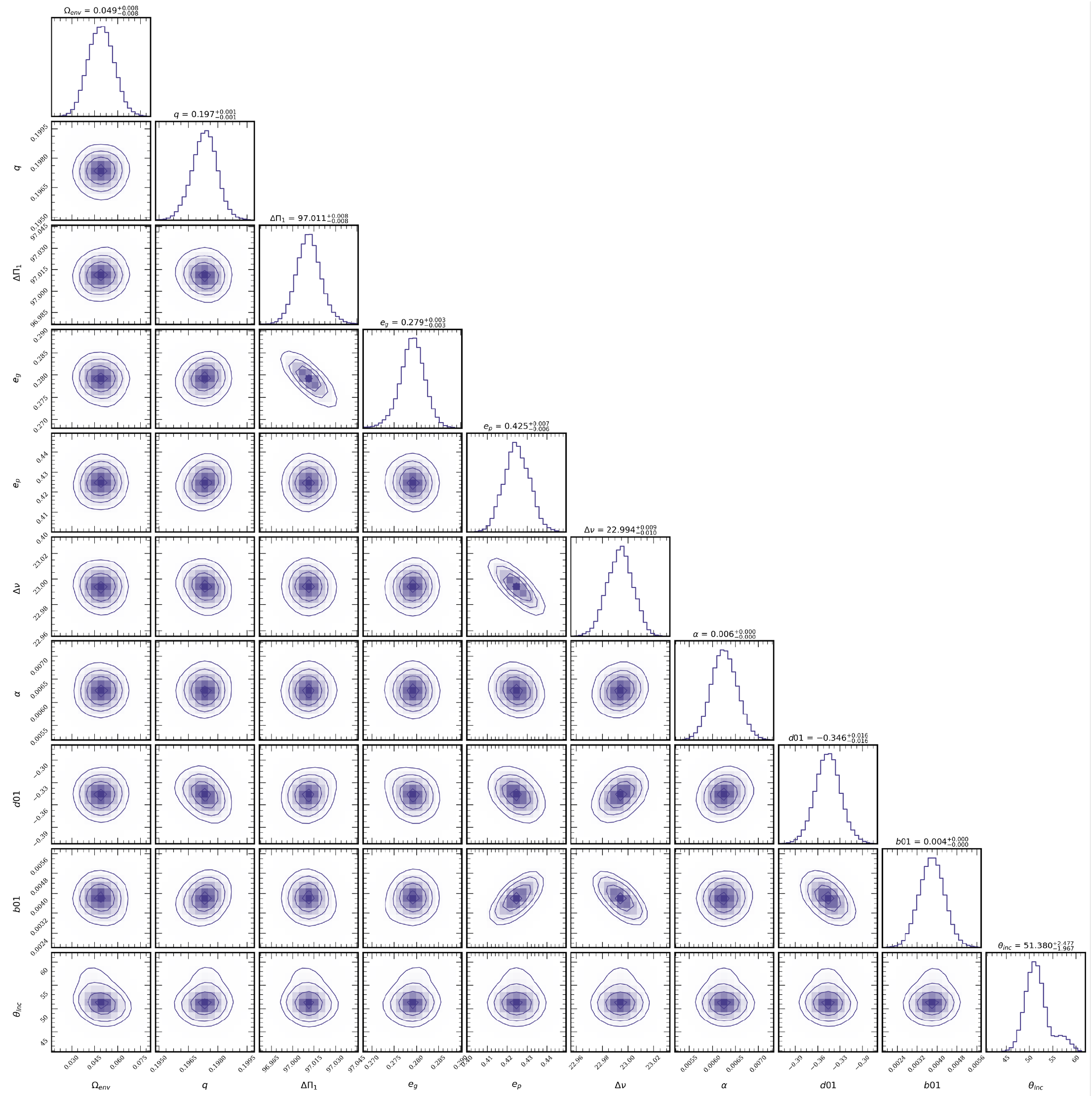}
    \caption{Cornerplot of the Bayesian fit of all the parameters for KIC 8179973 (second part).}
    \label{fig:corner_all_Alice_2}
\end{figure*}
\begin{figure*}
    \centering
    \includegraphics[width=\linewidth]{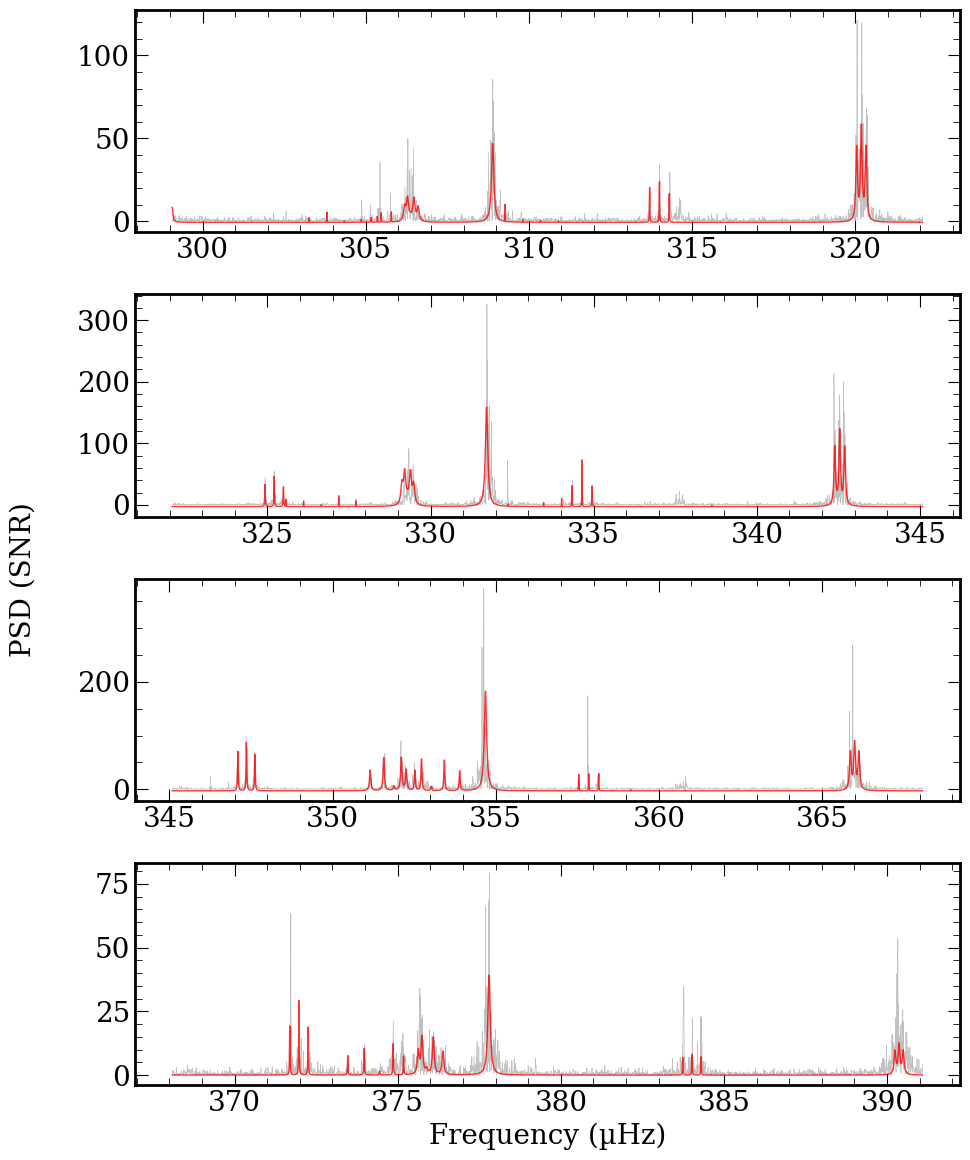}
    \caption{Echelle diagram showing the superimposed data and asymptotic model for the most prominent modes of KIC 8179973.}
    \label{fig:echelle_Alice}
\end{figure*}
\end{appendix}
\end{document}